\input harvmac
\input psfig
\newcount\figno
\figno=0
\def\fig#1#2#3{
\par\begingroup\parindent=0pt\leftskip=1cm\rightskip=1cm\parindent=0pt
\global\advance\figno by 1
\midinsert
\epsfxsize=#3
\centerline{\epsfbox{#2}}
\vskip 12pt
{\bf Fig. \the\figno:} #1\par
\endinsert\endgroup\par
}
\def\H{{\cal H}}
\def\figlabel#1{\xdef#1{\the\figno}}
\def\encadremath#1{\vbox{\hrule\hbox{\vrule\kern8pt\vbox{\kern8pt
\hbox{$\displaystyle #1$}\kern8pt}
\kern8pt\vrule}\hrule}}
\def\underarrow#1{\vbox{\ialign{##\crcr$\hfil\displaystyle
 {#1}\hfil$\crcr\noalign{\kern1pt\nointerlineskip}$\longrightarrow$\crcr}}}
%
\overfullrule=0pt
\def\bar{\overline}
%
\def\tilde{\widetilde}
\def\bar{\overline}
\def\hat{\widehat}
\def\Z{{\bf Z}}
\def\N{{\cal N}}
\def\T{{\bf T}}
\def\S{{\bf S}}
\def\R{{\bf R}}

\font\zfont = cmss10 
\font\litfont = cmr6

\def\bigone{\hbox{1\kern -.23em {\rm l}}}
\def\ZZ{\hbox{\zfont Z\kern-.4emZ}}
\def\half{{\litfont {1 \over 2}}}

\Title{hep-th/0006010}
{\vbox{
\centerline{ Supersymmetric
 Index}
\bigskip
\centerline{ In Four-Dimensional Gauge Theories}}}
\smallskip
\centerline{Edward Witten$^*$}
\smallskip
\centerline{\it Department of Physics, Caltech, Pasadena CA 91125}
\centerline{\it and}
\centerline{\it  CIT-USC Center for Theoretical Physics, Univ.
of Southern California, Los Angeles CA }\bigskip

\medskip

\noindent

This paper is devoted to a systematic discussion of the supersymmetric
index $\Tr\,(-1)^F$ for the minimal supersymmetric Yang-Mills theory -- 
with any simple gauge group $G$ -- primarily in four spacetime dimensions. 
The index has refinements that probe confinement and oblique confinement
and the possible spontaneous breaking of chiral symmetry 
and of global symmetries, such as
charge conjugation, that are derived from outer automorphisms of the gauge group.   Predictions for the index and its refinements are obtained on 
the basis of standard hypotheses about the infrared behavior of gauge 
theories.  The predictions are confirmed via microscopic calculations
which involve a Born-Oppenheimer computation of the spectrum as well as
mathematical formulas involving
triples of commuting elements of $G$ and the Chern-Simons
invariants of flat bundles on the three-torus.

\vskip 1cm
\noindent
$^*$ On leave from Institute for Advanced Study, Olden Lane,
Princeton, NJ 08540.

\Date{May, 2000}

\newsec{Introduction}

\def\Hom{{\rm Hom}}
A  constraint on the dynamics of supersymmetric field theories
is provided by the supersymmetric index $\Tr\,(-1)^F$ \ref\witten{E. Witten,
``Constraints On Supersymmetry Breaking,''  Nucl. Phys.
{\bf B202} (1982) 253.}.  It is defined
as follows.  One formulates an $n$-dimensional
supersymmetric theory of interest
on a manifold $\T^{n-1}\times \R$, where $\R$ parametrizes
 the ``time'' direction
and $\T^{n-1}$ is an $(n-1)$-torus (endowed with a flat metric and with 
a spin structure
that is invariant under supersymmetry,  that is with periodic boundary 
conditions for fermions in all directions).
If the original theory is such that the classical energy grows as one goes
to infinity in field space in any direction,\foot{This condition
notably excludes 
theories that have a noncompact flat direction in the classical
potential.} then the spectrum of the theory in a finite volume
 is discrete.  Under such conditions, let
$n_B$ and $n_F$ be the number of supersymmetric states of zero energy that
are bosonic or fermionic, respectively.  The supersymmetric index is defined
to be $n_B-n_F$ and is usually written as $\Tr\,(-1)^F$, where $(-1)^F$
is the operator that equals $+1$ or $-1$ for bosonic or fermionic states and
the trace is taken in the space of zero energy states.
Alternatively, as states of nonzero energy are paired (with equally many
bosons and fermions), one can define the index as
\eqn\goind{\Tr\,(-1)^Fe^{-\beta H},}
where $H$ is the Hamiltonian, $\beta$ is any positive real number, and the
trace is now taken in the full quantum Hilbert space.
This definition of the index shows that it can be computed as a path integral
on an $n$-torus $\T^n$, with a positive signature metric
of circumference $\beta$ in the time direction
(and periodic boundary conditions on fermions in the time direction
-- to reproduce the $(-1)^F$ in the trace -- as well as in the space 
directions).

The index is invariant under any smooth deformations of a supersymmetric
theory that leave fixed the behavior of the potential at infinity
(so that supersymmetric vacua cannot move to or from infinity).
The reason is that since the states of nonzero energy are paired,
any change in $n_B$ -- resulting from a state moving to or from zero energy
-- is accompanied by an equal change in $n_F$.
Because of this, it is usually possible to effectively compute the
index by perturbing to some sufficiently simple situation while preserving
supersymmetry.  A nonzero index implies that the ground state energy
is exactly zero for any volume of the spatial torus $\T^{n-1}$, and hence
that the ground state energy per unit volume vanishes in the infinite volume
limit.  It follows that if the infinite volume theory has a stable
ground state (which will be so if the potential grows at infinity) then
that state is supersymmetric.

This way of thinking illuminates supersymmetric
dynamics in many situations.  But application to four-dimensional
gauge theories was  hampered for some years
by the fact that, though one could readily
obtain attractive results for $SU(n)$ and $Sp(n)$ gauge groups,
the results for other groups appeared to clash with expectations
based on chiral symmetry breaking.  Ultimately (see the appendix
to \ref\vectorstructure{E. Witten, ``Toroidal Compactification Without
Vector Structure,''   hep-th/9712028.}), it became clear that the discrepancy
arose from overlooking the  fact that certain moduli spaces of
triples of commuting elements of a Lie group are not connected.

The purpose of the present paper is to 
systematically develop the theory of the supersymmetric index for
the minimal supersymmetric gauge theories in $2+1$ and $3+1$ dimensions.
Actually, for  the $(2+1)$-dimensional theories, we only
consider the special case that the Chern-Simons coupling is $k=\pm h/2$,
where confinement and a mass gap are expected.  The index in the
 $(2+1)$-dimensional
theories with more general Chern-Simons coupling  has been analyzed
elsewhere \ref\thrindex{E. Witten, ``Supersymmetric Index Of
Three-Dimensional Gauge Theory,'' hep-th/9903005.}.

In either $2+1$ or $3+1$ dimensions,
we consider supersymmetric theories with the smallest possible
number of supercharges (two or four in $2+1$ or $3+1$ dimensions)
and with the minimal field content: only the gauge fields and their
supersymmetric partners.   We will compute the index for bundles
of any given topological type on $\T^2$ or $\T^3$.  Because Yang-Mills
theory of a product group $G'\times G''$ is locally the product
of the $G'$ theory and the $G''$ theory, the index for a semisimple
$G$ can be inferred from the index for simple $G$ (provided that in the
simple case one has results for  all possible $G$-bundles).  
So we will assume that $G$ is simple.  

The index has a number of important refinements (some of which were treated
in \witten) that we will explain.
It is possible, by letting $G$ be non-simply-connected,
to include a discrete electric or magnetic flux and thereby probe
the hypothesis of confinement.  In four dimensions, it is possible
to refine the index to take into account a discrete chiral symmetry
group and to give evidence that the chiral symmetry is spontaneously
broken.  It is also possible, by allowing $G$ to be disconnected,
to give evidence that discrete symmetries associated with outer
automorphisms of $G$ (such as charge conjugation for $G=SU(n)$) are unbroken.

In sections 2 and 3, we explain what predictions about the supersymmetric 
index and its refinements
follow from standard claims about gauge dynamics.  
In section 4, we compute the index and its refinements 
microscopically, following the strategy of \witten\
but (in four dimensions)
including the contributions of all components of the moduli space
of flat connections.  Full agreement is obtained in all cases; in 
$3+1$ dimensions, obtaining such agreement 
depends on a result counting the moduli spaces of
commuting triples that was proposed in \vectorstructure\ and has
subsequently been justified.
For connected and simply-connected $G$, the moduli spaces of commuting
triples have been analyzed in 
\nref\keur{A. Keurentjes, A. Rosly, and A. V. Smilga, ``Isolated
Vacua In Supersymmetric Yang-Mills Theories,'' hep-th/9805183.}%
\nref\newkeur{
A. Keurentjes, 
``Non-Trivial Flat 
Connections On The 
3-Torus I: $G_2$ and the Orthogonal Groups,'' hep-th/9901154;
``Non-Trivial Flat Connections On The 3-Torus, II,''
hep-th/9902186.}%
\nref\kac{V. G. Kac and A. V. Smilga, ``Vacuum Structure In Supersymmetric
Yang-Mills Theory With Any Gauge Group,'' hep-th/9902029.}%
\nref\bfm{A. Borel, R. Freedman, and J. Morgan, ``Almost Commuting
Elements In Compact Lie Groups,'' preprint (1999). }%
\refs{\keur,\newkeur,\kac,\bfm}.  
A generalization to non-simply-connected $G$, necessary for including
the discrete fluxes, and an analysis of the Chern-Simons invariants
of flat bundles on $\T^3$, necessary for testing the claims about chiral
symmetry breaking, have been made in \bfm.
In section 5, we give some details about moduli spaces of commuting
triples.

\newsec{Expectations In $2+1$ Dimensions}

\subsec{Preliminaries}

 The minimal $(2+1)$-dimensional supersymmetric theory
has a field content consisting of the gauge field $A$ of some compact
connected
simple Lie group $G$, along with a Majorana fermion $\lambda$ in the
adjoint representation of the gauge group.  The usual kinetic energy
for these fields is
\eqn\nugo{L={1\over g^2}\int d^3x \,\Tr\left({1\over 4}
F_{ij}F^{ij}+{1\over 2}\bar\lambda i\Gamma\cdot D\lambda\right).}
However, there is  a crucial subtlety in $2+1$ dimensions: it is possible
to add an additional Chern-Simons coupling while preserving supersymmetry.
The additional interaction is
\eqn\newnugo{-{ik\over 4\pi}
\int\Tr\left(A\wedge dA +{2\over 3}A\wedge A\wedge A+\bar\lambda
\lambda\right).}
Here $k$ must, for topological reasons, obey a quantization condition.
For $G$ simply-connected, $k$ must be congruent to $h/2$ mod 1,
where $h$ is the dual Coxeter number of $G$; if $G$ is not simply-connected,
$k$ must be congruent to $h/2$ mod $s$ where $s$ is an integer
that depends on $G$.  The $h/2$ term in the quantization law comes from
an anomaly involving the fermions 
\nref\kao{H.-C. Kao, Kimyeong Lee, and Taelin Lee,
``The Chern-Simons Coefficient in Supersymmetric Yang-Mills
Chern-Simons Theories,'' hep-th/9506170.}
\refs{\kao,\thrindex}.

The supersymmetric index for general $k$ has been studied in \thrindex.
Our intent here is to describe what properties of the index, and related
invariants, can be deduced if we assume that in bulk the theory
has a unique vacuum with a mass gap and confinement.  As has been explained
in \thrindex, it is reasonable to believe that these properties
hold precisely if $k=\pm h/2$.  Hence in the present paper, when
we consider the $(2+1)$-dimensional theory, we will always specialize
to these values of $k$.

First we consider the theory with a gauge group $G$ that is simply-connected.
Since we assume a mass gap, there is no Goldstone fermion,
so that the unique vacuum has unbroken supersymmetry.  One expects that
an isolated vacuum with mass gap will contribute $+1$ to the supersymmetric
index, so one expects $\Tr\,(-1)^F=1$.
\foot{There is actually a small
subtlety here.  In $2+1$ dimensions, after compactifying on
a torus, there is no natural definition of the sign of the operator
$(-1)^F$, and hence the contribution of a massive vacuum to the
index might be $+1$ or $-1$.  It was shown
in \thrindex\ that with a natural convention, if the index is $+1$
for $k=h/2$, then it is $(-1)^r$ at $k=-h/2$, where $r$ is the rank of 
$G$.  For our purposes, we work at, say, $k=h/2$ and
define the sign of $(-1)^F$ so that the index is $+1$.}
The logic in this statement
 is that one can compute $\Tr\,(-1)^F$ by formulating the
theory on a two-torus of very large
radius $R$ with $1/R$ much smaller than the
mass gap; then on the length scale $R$, the system is 
locked in its ground state, which contributes $1$ to the index.
The mass gap is important here; an
isolated vacuum without a mass gap makes a contribution to the index
that is not necessarily equal to 1 (or even $\pm 1$), 
as was seen in some examples in \witten.

Now we consider the case that $G$ is not simply-connected.  Its fundamental
group $\pi_1(G)$ is necessarily a finite abelian group (cyclic
except in the special case $G=SO(4n)/\Z_2$, for which $\pi_1(G)=
\Z_2\times \Z_2$).  From a Hamiltonian point of view,
in quantizing on a two-torus $\T^2$, there are basically two changes
that occur when $G$ is not simply-connected.  First of all, a $G$-bundle
over $\T^2$ can be topologically non-trivial if $G$ is not simply-connected.
In the present paper, it will be important that if $X$ is a two-manifold
or a three-manifold, the possible $G$-bundles are classified by
a ``discrete magnetic flux,'' a characteristic class $m$ of the $G$-bundle
which takes values in  \eqn\xxy{M=H^2(X,\pi_1(G)).} 
(On a manifold of dimension
higher than three, a $G$-bundle has the characteristic class $m$ plus
additional invariants such as instanton number.) All values of $m$ can occur.
To gain as much information as possible, we do not want to sum over $m$;
we want to compute the index as a function of $m$.

\def\CW{{\cal W}}
The second basic consequence of $G$ not being simply-connected is that
the  group of gauge transformations, 
that is the group of maps of $\T^2$ to $G$ (or more
generally the group of bundle automorphisms if the $G$-bundle is non-trivial)
has different components.  
Restricted to a non-contractible loop in $\T^2$, a gauge transformation
determines an element of $\pi_1(G)$ which may or may not be trivial.
By restricting it to cycles generating $\pi_1(\T^2)$, a gauge
transformation $g$ determines a homomorphism $\gamma_g$ from
$\pi_1(\T^2)$ to $\pi_1(G)$; $g$ is continuously connected to the identity
if and only if $\gamma_g=0$.

Let   $\CW_0$ be the group of gauge transformations which are continuously
connected to the identity, and $\CW$ the group of all gauge transformations.
In quantizing a gauge theory, we must impose Gauss's law, which quantum
mechanically asserts that physical states must be invariant under $\CW_0$.
We need not impose invariance under $\CW$, and it will soon be apparent
that we can obtain more information if we do not impose such invariance.

\def\CH{{\cal H}}
The quotient
\eqn\duffo{\Gamma=\CW/\CW_0}
can be identified as follows:
\eqn\nuffo{\Gamma=\Hom(\pi_1(\T^2),\pi_1(G))=H^1(\T^2,\pi_1(G)).}
In particular, $\Gamma$ is a finite abelian group.  We define
the physical Hilbert space ${\cal H}_m$ 
by quantizing the space of connections on a bundle having characteristic
class $m$ and 
requiring invariance under $\CW_0$
only.  Then $\Gamma$ acts on $\CH_m$.  We can decompose ${\cal H}_m$ in
characters of $\Gamma$, in fact
\eqn\chardec{\CH_m=\oplus_e\CH_{e,m},}
where $\CH_{e,m}$ is the subspace of $\CH_m$ transforming 
in the character $e$ of $\Gamma$.  The character  $e$ is usually
called a ``discrete electric flux.''  It takes values in
\eqn\noffo{E=\Hom(\Gamma,U(1))=\Hom(H^1(\T^2,\pi_1(G)),U(1)).}

\bigskip
\centerline{\vbox{\hsize=4in\tenpoint
\centerline{\psfig{figure=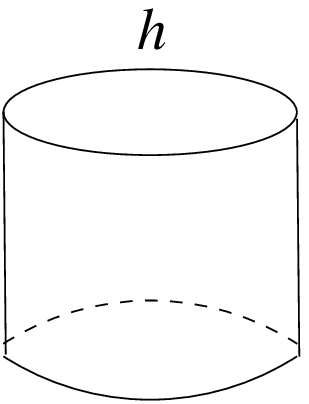}}
\vglue.1in
Fig. 1.  A nontrivial bundle on $\T^2$ can be built from a trivial
bundle on a cylinder by gluing together the two ends, using a gluing function
$h$.  The magnetic flux of the bundle is determined by the homotopy
class of $h$ in $\pi_1(G)$.}}

\bigskip\noindent{\it Constructions Of Non-Trivial Bundles}

For future reference, it will help to know that the two facts just
mentioned are closely related.  One way to make a non-trivial bundle
on $\T^2$ is to begin (figure 1) with a bundle on a cylinder $\S^1\times I$
(here $I=[0,1]$ is a unit interval) and glue the two ends together via
a gauge transformation.  The bundle on $\S^1\times I$ is inevitably
trivial ($\S^1\times I$ is contractible to $\S^1$, and
since $G$ is connected, a $G$-bundle on $\S^1$ is trivial).  
But in gluing the two ends together,
one can identify the fiber over $\S^1\times \{0\}$ with the fiber
over $\S^1\times \{1\}$ via an arbitrary gauge transformation, that
is an arbitrary map of $\S^1$ to $G$.   Up to homotopy, the map of
$\S^1$ to $G$ is determined by an element $f\in \pi_1(G)$, and for
every such $f$, we get a non-trivial $G$-bundle on $\T^2$.  Its
characteristic class is an element of $ H^2(\T^2,\pi_1(G))$ that we will
call $m(f)$.
Note in fact that, as there is essentially only one two-cycle on $\T^2$, namely
$\T^2$ itself,  $H^2(\T^2,\pi_1(G))$ is naturally isomorphic to $\pi_1(G)$;
the map from $f$ to $m(f)$ is an isomorphism.  $f$, if lifted to the
universal cover $\hat G$ of $G$, can be regarded as a path from the identity
in $\hat G$ to an element $\hat f$ of the center of $\hat G$.

A second and related construction
 is as follows.  Delete a point $P$ from $\T^2$.  As
$\pi_0(G)=0$, a $G$-bundle $W$ is trivial on the one-skeleton of any manifold
and hence on $\T^2-P$.  Likewise it is trivial in a small neighborhood
$U$ of $P$.  Let  $V$ be the intersection $V=(\T^2-P) \cap U$.
Comparing the two trivializations on $V$ gives a ``transition function''
$f:V\to G$.  As $V$ is homotopic to a circle, this gives an element of
$\pi_1(G)$, and thus, again, an element $\hat f$ of the center of $\hat G$.

As a variant of this,  
$W$, being trivial on either  $\T^2-P$ or $U$, can be lifted to a
 $\hat G$-bundle $\hat W$
on either $\T^2-P$ or $U$.  However, the two
lifts cannot be glued together to make a $\hat G$-bundle over $\T^2$;
the lifted transition function  is not single-valued,
but is multiplied by an element $\hat f$ of the center of $\hat G$
 in going around the circle.

In  this paper, a third  interpretation of  $G$-bundles on a torus
will also be helpful.  Every $G$-bundle on $\T^2$ admits a flat connection.
\foot{Pick a complex structure on $\T^2$.
Any connection on a $G$-bundle $W\to \T^2$ defines a holomorphic
structure on the complexification of $W$; this holomorphic structure is
semi-stable if the connection is chosen generically.  Such a semi-stable
holomorphic bundle  over a Riemann surface admits a flat
unitary connection by a theorem of Narasimhan and Seshadri; this
can be interpreted as a flat connection on the original $G$-bundle.}
A flat connection has holonomies around the two directions
in $\T^2=\S^1\times \S^1$; the holonomies are two elements of $G$,
say $a$ and $b$, which commute.
Suppose, however, that one lifts $a$ and $b$ to elements, say
$A$ and $B$, in the universal cover $\hat G$ of $G$.
There is no natural way to do this; pick any choice.    Then 
\eqn\juffy{\hat f=ABA^{-1}B^{-1}}
is an element of $\hat G$
that projects to the identity in $G$.  Thus, $\hat f$ is in particular
an element of the center of $\hat G$.
($\hat f$ is independent of the choice of lifting of $a$ and $b$,
since different liftings differ by multiplication of $A $ and $B$
by central elements 
of $\hat G$ which cancel out in the definition of $\hat f$.)

We thus have three different ways to associate a flat $ G$-bundle over
$\T^2$ with an element $f\in \pi_1(G)$ or $\hat f$ in the center of $\hat G$:

(1) Pull the bundle back to a flat bundle on a cylinder, as in figure 1,
and identify the gluing data with $f$.

(2) Trivialize the bundle away from and near a point $P\in \T^2$,
and find the $f$ that appears in comparing the two trivializations.

(3) Find a flat connection and measure $\hat f$ as in \juffy.  

It can be shown that the three approaches are equivalent.

Following \bfm, we will call group elements $A,B\in\hat G$ that commute
if projected to a quotient $G$ of $\hat G$ (and hence $ABA^{-1}B^{-1}$ is
in the center  of $\hat G$) ``almost commuting.''

\subsec{ The Index}

Supersymmetry acts in each $\CH_{e,m}$, so we can define the index
$\Tr\,(-1)^F$ for each value of $e$ and $m$; we call this index function
$I(e,m)$.  Our goal, in the $(2+1)$-dimensional case, is to predict
$I(e,m)$ based on the usual hypotheses about the infrared dynamics of this
theory.  In section 4.1, we will compare the predictions to a microscopic computation.
Of course, the predictions will only hold for those values of 
$k$ (namely $k=\pm h/2$) at which the theory has confinement, a mass gap,
and a unique vacuum.  The reason for analyzing this case in so much
detail is largely that it is good background for $3+1$ dimensions.

I will first explain why one intuitively expects $I(e,0)=0$ for $e\not= 0$
 on the basis of
confinement.  We let $|\Omega\rangle$ be the supersymmetric
ground state in $\CH_{0}$. 
In lattice gauge theory (which is difficult to implement technically
for the supersymmetric theory, but should give us a guide about the meaning
of confinement), one associates $|\Omega\rangle$ with a state that
in the strong
coupling limit 
is independent of the connection $A$.
Such a state is certainly $\Gamma$-invariant.
 Thus, $|\Omega\rangle$ should in particular be
a $\Gamma$-invariant state.
Hence, when we decompose $\CH_0=\oplus_e\CH_{e,0}$, the supersymmetric
state appears for $e=0$, so that $I(0,0)=1$, $I(e,0)=0$ for $e\not=\ 0$.

To see what is going on more explicitly, we will describe a typical
state in $\CH_{e,0}$ for $e\not= 0$.
Let $C$ be a noncontractible loop
in $\T^2$ that wraps once around one of the two circles in $\T^2=\S^1\times
\S^1$.  We let $V$ be an irreducible representation 
of the simply-connected universal
cover $\hat G$ of $G$ that is not a representation of $G$.  Consider the Wilson
loop operator around $C$ in the $V$ representation:
\eqn\noffo{W_V(C)=\Tr_VP\exp\oint_CA.}
It is invariant under $\CW_0$, but not under $\CW$; it transforms in
some nontrivial character $e$ of $\Gamma$. ($e$ is determined by the action
of the center of $\hat G$ on $V$.)  Consider the state
\eqn\poffo{W_V(C)|\Omega\rangle.}
It lies in $\CH_{e,0}$.  Now, intuitively confinement means\foot{This is an
oversimplification, for the classification of massive phases of gauge
theories is          more complicated than traditionally supposed,
as shown in \nref\kreskill{L. Krauss and J.  Preskill,
``Local Discrete Symmetry And Quantum Mechanical Hair,'' Nucl. Phys.
{\bf B341} (1990) 50.}%
\refs{\kreskill,\thrindex},
but the statement
 holds in the simplest
universality class of confining theories.  We will test the hypothesis
that the theories under discussion are in this universality class.}  
that the energy of the state \poffo, or any state carrying the same
electric flux, should grow linearly with the radius $R$ of $\T^2$.
In particular, in the space $\CH_{e,0}$, there should be no zero energy
supersymmetric state if $R$ is large enough, 
so $I(e,0)$ should vanish if the theory is confining.

Now let us reformulate the vanishing of $I(e,0)$
 in terms of path integrals (essentially
following an argument due to 't Hooft \ref\thooft{G. 't Hooft,
``On The Phase Transition Towards Permanent Quark Confinement,''
Nucl. Phys. {\bf B138} (1978) 1, ``A Property Of Electric And Magnetic
Flux In Non-Abelian Gauge Theory,'' Nucl. Phys. {\bf B153} (1979) 141,
``Topology Of The Gauge Condition And New Confinement Phases In
Non-Abelian Gauge Theories,'' Nucl. Phys. {\bf B205} (1982) 1.}).  
Let $|\Gamma|$ denote the number of elements of $\Gamma$. 
The projection operator onto states that transform in the character
$e$ is
\eqn\hubbo{P_e={1\over |\Gamma|}\sum_{\gamma\in\Gamma}e^{-1}(\gamma)\,\gamma.}
We can therefore write
\eqn\imlp{\eqalign{
I(e,0)=&\Tr_{\CH_{e,0}}(-1)^Fe^{-\beta H}=\Tr_{\CH_{0}} P_e(-1)^F
e^{-\beta H}
\cr=&{1\over |\Gamma|}\sum_{\gamma\in\Gamma}e^{-1}(\gamma)\Tr_{\CH_{0}} \gamma 
(-1)^Fe^{-\beta H}.\cr}}
Here $\beta$ is an arbitrary positive real number.

Now, let us consider a computation of 
\eqn\nimlop{\Tr_{\CH_0}\gamma (-1)^Fe^{-\beta H}}
via path integrals.  For $\gamma=1$, this was already explained
in the introduction: the trace in question is represented by
a path integral on a three-torus $\T^3=\T^2\times \S^1$, where $\S^1$
is a circle of circumference $\beta$ and the boundary conditions
for the fermions are periodic in all directions.  To insert $\gamma$ in the trace, we simply proceed as in figure 2.  We begin with $\T^2\times I$, where
$I$ is the closed interval $[0,\beta]$ of length $\beta$.  We glue
together $\T^2\times \{0\}$ with $\T^2\times \{\beta\}$ to make a trace.
But in the gluing, we identify the two ends via the gauge transformation
$\gamma$ (that is, via any gauge transformation that is in the homotopy class
of $\gamma$).  Such gluing is a standard way, introduced in connection
with figure 1, to construct a non-trivial $G$-bundle over $\T^3$.

\bigskip
\centerline{\vbox{\hsize=4in\tenpoint
\centerline{\psfig{figure=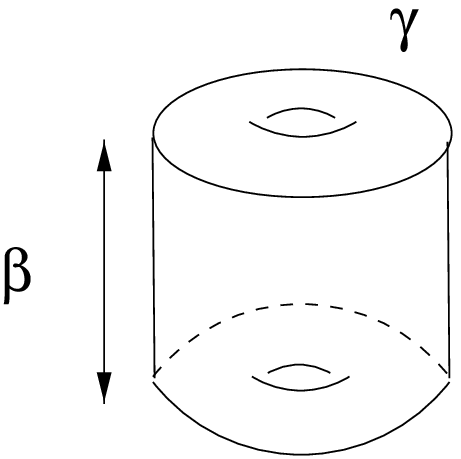}}
\vglue.4in
Fig. 2.  A  path integral computation of $\Tr_{\CH_0}\gamma (-1)^F e^{-\beta H}$
is made by performing a path integral on $\T^2\times \S^1$.  The $G$-bundle
on $\T^2\times \S^1$ is built from a $G$-bundle on $\T^2\times I$,
with $I$ a unit interval of length $\beta$, 
by using the gauge transformation $\gamma$
as a gluing function.
}}

Let $m_0(\gamma)$ be the characteristic class of the bundle obtained in this way.
To describe $m_0(\gamma)$ more fully,
let $x^1$ and $x^2$ be angular coordinates on the ``space'' manifold
$\T^2$, and $x^3$ an angular coordinate on $\S^1$.
Then $m_0(\gamma)\in H^2(\T^2\times\S^1,\pi_1(G))$ has vanishing
1-2 component, while the 1-3 and 2-3 components are determined by
$\gamma$.  For any $\hat m\in H^2(\T^2\times \S^1,\pi_1(G))$ we call
the 1-2 component ``magnetic'' and the 1-3 and 2-3 components
``electric.'' So $m_0(\gamma)$ is purely electric.  We will
write $\hat m$ for a general element of $H^2(\T^2\times \S^1,\pi_1(G))$,
and $m$ for its magnetic part, that is, the restriction of $\hat m$ to
$\T^2$.

For any $G$-bundle on $\T^3$ with characteristic class $\hat m$,
let us write $Z(\hat m)$ for the
 partition function on $\T^3$, with periodic boundary
conditions for fermions in all directions.  Then from the discussion
of \nimlop,
\eqn\olip{\Tr_{\CH_0}\gamma (-1)^Fe^{-\beta H}=Z(m_0(\gamma))}
and hence from \imlp\ we get
\eqn\ikl{I(e,0)={1\over |\Gamma|}\sum_{\gamma\in\Gamma}e^{-1}(\gamma)
Z(m_0(\gamma)).}

Now, $m_0(\gamma)$ is not a completely general element of $H^2(\T^2\times \S^1,
\pi_1(G))$, because its magnetic part, that is the restriction
to $\T^2$, vanishes.
This is because  in our above construction,
we started with a trivial bundle on the original $\T^2$.
But $m_0(\gamma)$ is subject to no other restriction.
A general $\hat m\in H^2(\T^2\times \S^1,\pi_1(G))$ has
components $\hat m_{12}$, $\hat m_{13}$, and $\hat m_{23}$.
Vanishing of the magnetic part means that $\hat m_{12}=0$.

Next, as discussed above, we interpret confinement to mean that $I(e,0)=0$
for all non-zero $e$.  Then \ikl\ implies that $Z(m_0(\gamma))$ is
independent of $\gamma$.  Since $Z(m_0(0))=Z(0)$ is the original
index, which is expected to equal 1 since the theory in infinite volume
has a unique vacuum with mass gap, we get
\eqn\olfo{Z(\hat m)=1\,\,\,{\rm if}\,\,\hat m_{12}=0.}

\subsec{ Action Of $SL(3,\Z)$}

So far, on $\T^3=\T^2\times \S^1$, we have considered a flat metric
for which the two factors are orthogonal.  More generally, we could
consider any flat metric on $\T^3$; because of the interpretation of $Z(\hat m)$
as an index, it is independent of which flat metric we pick.
To justify the last statement in detail, note that the path integral on 
 $\T^3=\T^2\times \S^1$ with trivial spin structure and 
specified $\hat m$
can be interpreted as 
\eqn\jungly{\Tr_{{\cal H}_{m}}\gamma (-1)^F e^{-\beta H+i\vec
a\cdot \vec P}.}
 Here notation is as follows: $m$ is the 1-2 component
of $\hat m$, which we regard as the characteristic class of a bundle on $\T^2$; 
$\H_{m}$ is the Hilbert space for quantization with this bundle; 
$\gamma$ is such that $m_0(\gamma)$ as defined earlier has the same
$13$ and $23$ components as $\hat m$;
$\vec P$ are the momentum operators; and $\beta$ and
$\vec a$ are constants that
contain the information needed to specify the $\T^3$ metric once the $\T^2$
metric is given.  ($\vec a=0$ if and only if the two factors in $\T^2\times
\S^1$ are orthogonal.)   As $H$ and $\vec
P$ commute with supersymmetry and vanish for supersymmetric states,
this trace can as usual be computed just in the space of supersymmetric states
and is independent of $\vec a$ and $\beta$; it similarly is independent
of the metric on $\T^2$ because supersymmetric states can only appear
or disappear in bose-fermi pairs when that metric is varied.  This
justifies the claim that $Z(\hat m)$ is independent of the flat metric on 
$\T^3$.\foot{This claim could alternatively be proved 
by noting that the derivative of $Z(\hat m)$ with respect to the metric
is given by a one-point function of the stress tensor $T_{mn}$.
But this one-point function vanishes because $T$ is of the form $\{Q,\dots\}$;
in fact	$T_{mn}=\gamma_m^{\alpha\beta}\{Q_\alpha,S_{ n\beta}\}+m\leftrightarrow
n,$
with $S_{n\beta}$ the supercurrent and $Q_\alpha $ the supercharges.}

$\T^3$ has a group of orientation-preserving diffeomorphisms
isomorphic to $SL(3,\Z)$, acting on the angles $x^1,x^2,x^3$ in the natural
way.  An $SL(3,\Z)$ transformation certainly leaves invariant the path integral
 if we allow for the action of $SL(3,\Z)$ on both the metric
of $\T^3$ and $\hat m$.  As $Z(\hat m)$ is independent of metric, it is 
invariant under the action of $SL(3,\Z)$ just on $\hat m$.

Since we are in three dimensions, a two-form is dual to a vector,
and we can identify $\hat m$ with a three-plet $\vec w=
(w_1,w_2,w_3)=(\hat m_{23},\hat m_{31},\hat m_{12})$
of elements of $\pi_1(G)$. $SL(3,\Z)$ acts on $\vec w$ in the natural
three-dimensional representation, tensored with $\pi_1(G)$.  
If
$\pi_1(G)$ is a cyclic group (isomorphic to $\Z_n$ for some $n$),
then $\vec w$ is just a single vector with coefficients in $\Z_n$.
Let $\mu$ be the greatest common divisor of the $w_i$ and $n$
(an integer prime to $n$ is invertible mod $n$, so only divisors of the
$w_i$ that divide  $n$ are meaningful).  One can lift
$\vec w/\mu$ to a vector $\vec w'$ with integer coefficients that
is ``primitive'' (the components are relatively prime).
One can find a basis of the lattice in which $\vec w'$ is the first vector;
this basis is related by an $SL(3,\Z)$ transformation to the original
basis.  This $SL(3,\Z)$ transformation, which we have constructed
      assuming            $\pi_1(G)$ is cyclic, puts                                      $\vec w$ in the form
$\vec w=(\mu,0,0)$; in particular, 
this shows that when $\pi_1(G)$ is cyclic,
the greatest common divisor $\mu $ is
the only $SL(3,\Z)$ invariant of $\hat m$.  $\vec w=(\mu,0,0)$ means that
$\hat m_{12}=\hat m_{31}=0$.

Actually,
$\pi_1(G)$ is cyclic for all simple $G$ unless $G=SO(4n)/\Z_2$,
in which case $\pi_1(G)=\Z_2\times \Z_2$.  For this group,
 $\vec w$ consists of a pair
of $\Z_2$-valued vectors. (Since 2 is prime, we can assume in this case
that $\mu=1$.)  By an $SL(3,\Z)$ transformation we can put
the first vector in the form $(w,0,0)$, and then, using the $SL(2,\Z)$ that
leaves this vector invariant, we can put the second, if not zero, in the form
  $(w',w'',0)$.
In particular,
we can set $\hat m_{12}=0$.  In short, in all cases we can set $\hat m_{12}=0$
by an $SL(3,\Z)$ transformation.

We can thus extend \olfo:
\eqn\bolfo{Z(\hat m)=1\,\,{\rm for~all}\,\,\hat m\,\,{\rm and}\,\,G.}

\bigskip\noindent{\it General Prediction For The Index}

Now we can state the general prediction for the index in three spacetime
dimensions.

First consider $I(0,m)$.  This is the same as the path integral
 $Z(\hat m)$ with $\hat m_{12}=m$, $\hat
m_{13}=\hat m_{23}=0$.  Hence,
by virtue of \bolfo\ we get
\eqn\kiko{I(0,m)=1\,\,{\rm for~all}\,\,m.}

To compute $I(e,m)$ in general, we 
introduce some notation.  Given $m\in H^2(\T^2,\pi_1(G))$,
and $\gamma\in \Hom(\pi_1(\T^2),\pi_1(G))$, we write $\hat m( m,\gamma)$
for the element of $H^2(\T^2\times \S^1,\pi_1(G))$ such that $\hat
m_{12}=m$ (that is, the restriction of $\hat m$ to $\T^2$ equals $m$)
while the 1-3 and 2-3 components of $\hat m( m,\gamma)$ 
equal those of $m_0(\gamma)$ as defined above.
Then arguing as we did in connection
with figure 2 but starting with a bundle whose characteristic class on
$\T^2$ is $     m$, we get
\eqn\ikli{I(e,m)={1\over |\Gamma|}\sum_{\gamma\in\Gamma}e^{-1}(\gamma)
Z(\hat m( m,\gamma)).}
  Using \bolfo, we therefore get
\eqn\nikli{I(e,m)=0~{\rm for}~e\not=0}
for all $m$ and $G$.

\kiko\ and \nikli\ are the basic predictions concerning the index in three
dimensions.  In section 4.1,
 we will confront them with microscopic computations.
For now, we move on to a discussion of the analogous predictions in
four dimensions.

\newsec{The Four-Dimensional Case}

\subsec{Simply-Connected Gauge Group}

We will now consider the minimal $(3+1)$-dimensional super Yang-Mills
theory in a similar way.  Again we start with the case that the gauge
group $G$ is simply-connected.  The Lagrangian is
\eqn\onugo{L={1\over g^2}\int d^3x \,\Tr\left({1\over 4}
F_{ij}F^{ij}+{1\over 2}\bar\lambda i\Gamma\cdot D\lambda\right)+{\theta\over
8\pi^2} \int \,\Tr F\wedge F.}
The last term is a ``topological'' term that has no close analog
in $2+1$ dimensions, while on the other hand the $(2+1)$-dimensional
Chern-Simons term has no analog in $3+1$ dimensions.
Also $\lambda$ was real in $2+1$ dimensions, but 
in $3+1$ dimensions, $\lambda$ is a positive chirality fermion in the adjoint
representation of $G$, and $\bar\lambda$ is its complex conjugate.

That last
remark leads at once to an important difference between the $2+1$ and
$3+1$-dimensional problems.  In $3+1$ dimensions, we have at the
classical level the $U(1)$ chiral symmetry
\eqn\kiop{\lambda\to e^{-i\alpha}\lambda,~~\bar\lambda\to e^{i\alpha}
\bar\lambda,}
with $\alpha$ an angular parameter.  
We describe this transformation law by saying that $\lambda$ and
$\bar\lambda$ have respectively charge $-1$ and charge $1$ under the symmetry.
Quantum mechanically, $U(1)$
is broken by an anomaly to a subgroup $\Z_{2h}$, where $h$ is the dual
Coxeter number of $G$.  
This is computed by showing that in an instanton field, the index
of the Dirac operator for the $\lambda$ field is
\eqn\furgy{{\rm ind}(\lambda)=2h.}
The surviving $\Z_{2h}$ is an exact symmetry group of the quantum
theory.  It contains a subgroup $\Z_2$,
which is generated by a group element that
multiplies both $\lambda$ and $\bar\lambda$ by $-1$.  This element
is equivalent to a $2\pi$ rotation in spacetime and thus cannot be
spontaneously broken.  
  By analogy with chiral symmetry breaking in the
strong interactions, it is believed that if the theory is formulated
on flat $\R^4$, the $\Z_{2h}$ symmetry is spontaneously
broken to $\Z_2$; this is its maximal subgroup that would allow a bare
mass for $\lambda$.  If so, there are at least $h$ vacua.  It is believed
that there are precisely $h$ vacua and that these vacua all have a mass
gap and confinement.

If then the theory is formulated on $\T^3\times \R$, where $\T^3$ is
a very large three-torus and $\R$ parametrizes ``time,''
then 
these vacua (as they differ by a symmetry rotation) all make the same
contribution to the index, and this contribution
 is $\pm 1$ because of the mass
gap.  With one natural normalization of the sign of $(-1)^F$, which
we will describe in section 4, the sign is $(-1)^r$, where $r$ is the rank
of $G$, and the index is $\Tr\,(-1)^F=(-1)^rh$.  
(For many purposes, one might redefine the sign of $(-1)^F$ to eliminate
the factor $(-1)^r$.)
However, in $3+1$ dimensions, we can
define an invariant more refined than the index.

Let $\H$ be the Hilbert space of  the theory on $\T^3$.  
Of course, the supersymmetry
algebra requires that states of zero energy also have zero three-momentum.
We can decompose $\H$ in the form
\eqn\mixx{\H=\oplus_{k=0}^{2h-1}\H_k,}
where $\H_k$ consists of  states that transform under the
chiral symmetry with charge $k$, that is
by
\eqn\nixop{\psi\to e^{ik\alpha}\psi.}
Since an anomaly breaks the chiral $U(1)$ to $\Z_{2h}$, $k$ is only
defined modulo $2h$, a fact that is incorporated in the notation in \mixx.
Since elementary fermi fields all have charge $\pm 1$ while elementary
bosons are invariant under the chiral symmetry, states in $\H_k$ are
all bosonic or all fermionic for even or odd $k$:
\eqn\kilo{(-1)^F=(-1)^k.}
Actually, we should be more careful here.  All we really know about
$(-1)^F$ and $(-1)^k$ is that they transform all local fields in the same
way.  This would allow a sign factor in the relation between these
two operators, but with the conventions we will use for the microscopic
computation in section 4, the sign is $+1$.

\def\HK{H^k(Q)}
\def\hk{h^k(Q)}
Let $Q_\alpha$
and $\bar Q_{\dot\alpha}$ be the supercharges of positive and negative
chirality (here $\alpha$ and $\dot\alpha$ are spinor indices of the two
chiralities).  Then $Q_\alpha $ and $\bar Q_{\dot\alpha}$ have respectively
charge $1$ and charge $-1$ under $\Z_{2h}$.  Let $ Q$ be any generic linear
combination of the $Q_{\alpha}$, and $ \bar Q$ its hermitian conjugate. 
Then the supersymmetry algebra implies that
\eqn\mico{ Q^2=0,}
and 
because $Q$ has charge 1, $Q$ maps $\H_k$ to $\H_{k+1}$.
In this situation, we can define the {\it cohomology groups} of $Q$:
\eqn\groky{H^k(Q)={\rm ker}(Q:\H_k\to \H_{k+1})/{\rm im}(Q:\H_{k-1}\to \H_k).}
Let $\hk$ be the dimension of $\HK$.

The supersymmetry
algebra implies that (after multiplying $Q$ by a suitable scalar)
\eqn\kico{\{\bar Q,Q\}=2(H+\vec a \cdot \vec P)}
with $H$ the Hamiltonian, $\vec P$ the momentum, and $\vec a$ a unit
three-vector that depends on the choice of $Q$. 
Given \kico, a standard ``Hodge theory'' argument
says that the $\HK$ consist  of  states in $\H_k$ annihilated by
$H+\vec a\cdot \vec P$.  If there is a mass gap, this means that the $H^k(Q)$ 
consist of the states annihilated by both $H$ and $\vec P$ (massive particles
have $H>|\vec P|$).\foot{Even if there is no mass gap, states annihilated
by $H+\vec a\cdot \vec P$ are annihilated separately by $H$ and $\vec P$
for generic $\vec a$ -- and hence for generic $Q$ -- as $H$ and $\vec P$
both have discrete spectra.}
This shows that $H^k(Q)$ is independent of $Q$.  It also implies,
given \kilo, that
\eqn\omigo{\Tr\,(-1)^F=\sum_k(-1)^{k}\hk.}

Let us compute the $\hk$ for the case of a large volume on $\T^3$.
Let $v$ be the generator of $\Z_{2h}$ obtained by setting
$\alpha =\pi/h$ in \kiop.  Let $|\Omega\rangle $ be any one supersymmetric
state of the system that obeys cluster decomposition in the infinite
volume limit, and let $|\Omega_s\rangle=v^s|\Omega\rangle$ for any
integer $s$.   Note that $|\Omega_{s+h}\rangle = (-1)^r|\Omega_s\rangle$,
where the sign arises because $v^h=(-1)^F$ and
all of these states have $(-1)^F=(-1)^r$.
The $|\Omega_s\rangle$ are thus cyclically permuted by the broken
symmetry, and spontaneous symmetry breaking implies that they are linearly
independent.  Let
\eqn\jurry{|\Theta_k\rangle=\sum_{s=0}^{h-1}e^{-\pi i k   s/h}
|\Omega_s\rangle}
for any integer $k$ such that $k+r$ is even.
$|\Theta_k\rangle$ thus has charge $k$.
The $|\Theta_k\rangle$ span the space of zero energy states, so
\eqn\ucases{\hk=\left\{\matrix{1,~{\rm for~even}~k+r\cr
0,~{\rm for~odd}~k+r.\cr}\right. }
This is clearly compatible with the previous claim that
$\Tr\,(-1)^F=(-1)^rh $.

In this theory, the chiral
supercharges $Q_\alpha$ vary holomorphically as a function of
\eqn\jico{\tau={\theta\over 2\pi }+{4\pi i\over g^2}.}
(The anti-chiral supercharges vary anti-holomorphically.)
$\theta$ and $g$ are the theta angle and coupling constant of the theory
defined with a specific renormalization scheme; 
one could use the renormalization
group to replace $g $ with a mass parameter used in the renormalization.
Because of the holomorphy, one
 expects values of $\tau$ at which there appear extra zero energy states, not
present generically, to be isolated points in $\tau$ space.   

The result obtained in \ucases\ follows from the existence of a mass gap
and holds  for fixed coupling and
any sufficiently large volume, or equivalently for fixed volume and
sufficiently large coupling.   Since it holds not just at an isolated
set of points, it must be the generic result.  This important
statement can
also be justified in either of the following two ways:

(1) Any ``jumping'' phenomenon consists of the appearance of extra zero
energy states  in bose-fermi pairs.\foot{In fact, 
jumping involves appearance of extra states with adjacent
values of $k$.
Any states that come to or from zero energy can be naturally grouped
in pairs of states related by the action of $Q$.
Since $Q$ has charge 1, 
such a pair consists of a boson of charge $k$ and a fermion of charge
$k+1$ or $k-1$, for some $k$.}     Continuity of the spectrum as a function
of coupling implies that zero energy states present generically
cannot disappear at a special value of the coupling.
The generic value of the $\hk$ is thus the minimum value that they
attain for any value of the coupling. 
Since according to \ucases, the $\hk$ vanish for odd $k$, this must
be the correct generic value for odd $k$, and hence also for even $k$.

(2) By arguments in \witten\ that involve showing that $Q$ changes
by conjugation under a change in volume or coupling, it can be shown
in this particular problem that there is actually no jumping for any $\tau$.

Since \ucases\ is the generic value, it can be compared to a weak
coupling computation peformed for sufficiently small $g$ and any $\theta$.
Such a comparison will be the goal of section 4.  

Before going on to the case that $G$ is not simply connected,
we recall the Hamiltonian interpretation of the theta angle
in gauge theories.  The group of gauge transformations
on $\T^3$ is not connected.   If $\pi_1(G)=0$, the components
are labeled by a ``winding number'' (or degree) associated with
\eqn\junglo{\pi_3(G)=\Z.}
Let $T$ be a gauge transformation of ``winding number one.''
Just as in the discussion of discrete electric flux in section 2,
in quantizing the theory we must consider states that are invariant
under the group ${\cal W}_0$ of gauge transformations that are continuously
connected to the identity; but we need not require invariance under $T$.
Rather,  we can select an arbitrary angle $\theta$ and require
that our quantum states $|\psi\rangle$ obey
\eqn\jangle{T|\psi\rangle =e^{i\theta}|\psi\rangle.}

\subsec{Non-Simply Connected Case}

If $G$ is not simply-connected (but, for
simplicity, still connected), then, as in $2+1$ dimensions,  
quantum states can be labeled by discrete magnetic
and electric charges.  As in \xxy\ and \noffo, the discrete magnetic
charge is a characteristic class of the $G$-bundle that takes values in
\eqn\kiddo{M=H^2(\T^3,\pi_1(G)),}
and the discrete electric charge is a quantum number that takes values in
\eqn\niddo{E=\Hom(H^1(\T^3,\pi_1(G)),U(1)).}
Actually, there is a subtlety already at this point, because on $\T^3$,
unlike $\T^2$, an element of $H^1(\T^3,\pi_1(G))$ does not uniquely
determine the homotopy class of a gauge transformation.  There is
also the winding number discussed in the last paragraph.  For the moment,
we will avoid this issue by assuming that $\theta=0$, so that the winding
number does not affect how a gauge transformation acts on physical states.
We restore the theta dependence in section 3.4
 when we analyze oblique confinement.

For some purposes, there is no essential loss to suppose that $G$, if
not simply connected, is the adjoint form of the group, which is the
form of the group with trivial center and the largest possible $\pi_1(G)$.
For this choice of $G$, we meet the widest possible set of values of $m$
and $e$.  Other choices of $G$ correspond to considering only a subset
of conceivable values of $m$ and $e$.

\bigskip\noindent{\it Identification Of $E$ And $M$}

Before discussing the predictions for the supersymmetric index
and the $h^k(Q)$'s, we will make a small digression (which the reader
may choose to omit).
$(3+1)$-dimensional gauge theories can sometimes have nonperturbative dualities
that exchange electric and magnetic fields.  We will not actually
in this paper study any example in which this happens.  But for
any such example to exist, it must be that in three dimensions
the spaces $M$ and $E$ are the same, so that it is possible to have
a nonperturbative duality exchanging $m$ and $e$.\foot{For instance,
in \ref\vw{C. Vafa and E. Witten, ``A Strong Coupling Test Of
$S$ Duality,'' Nucl. Phys. {\bf B431} (1994) 3, hep-th/9408074.}, 
it was shown that Montonen-Olive duality
for $N=4$ super Yang-Mills theory exchanges $m$ and $e$.}

First of all, by Poincar\'e duality, \niddo\ can be rewritten in the
form
\eqn\ubiddo{E=H^2(\T^3,\tilde{\pi_1(G)}),}
where 
\eqn\nubiddo{\tilde {\pi_1(G)}=\Hom(\pi_1(G),U(1 )).}
Now, if $\Gamma$ is any finite
abelian group, then $\tilde\Gamma=\Hom(\Gamma,U(1))$
is isomorphic to $\Gamma$.  So the groups $E$ and $M$ are isomorphic
as abstract abelian groups.  However, there is for general $\Gamma$
no {\it natural} isomorphism between $\Gamma$ and $\tilde \Gamma$
(generally no such isomorphism is invariant under all automorphisms
of $\Gamma$).
For nonperturbative dualities exchanging $e$ and $m$ to be possible,
it should be that there is a natural isomorphism between $E$ and $M$,
which will be so if there is a natural isomorphism between $\pi_1(G)$ 
and $\tilde{\pi_1( G)}$.   
This can be constructed as follows:\foot{The following
was pointed out by J. Morgan.}

(1) If $G$ is not simply-laced, then $\pi_1(G)$ is trivial or $\Z_2$,
and either way there is only one possible isomorphism between $\pi_1(G)$ and
$\tilde{\pi_1(G)}$. Being unique, this isomorphism is natural.

(2) In discussing the simply-laced case, we first assume
 that $G$ is the adjoint group, with trivial center.  Then if
$\Lambda$ is the root lattice of $G$ and $\Lambda^\vee$ (its dual)
is the weight lattice, we have $\pi_1(G)=\Lambda^\vee/\Lambda$.  Let
$(~,~)$ be the inner product on $\Lambda^\vee$ (normalized as usual so that roots
of $G$ have length squared two).  It has the property
that $(x,y)\in \Z$ for $x\in \Lambda^\vee$, $y\in \Lambda$, but not, in 
general, if $x,y\in\Lambda^\vee$.  Hence, if reduced mod 1, $(x,y)$ is
a well-defined pairing on $\Lambda^\vee/\Lambda=\pi_1(G)$.  This pairing
takes values in
${\bf Q}/{\bf Z}$, which we can regard as a subgroup of $\R/\Z=U(1)$.
This pairing lets us define a map from $\pi_1(G)$ to $\tilde {\pi_1(G)}$
by sending $x$ to the element $\phi_x\in \tilde {\pi_1(G)}$ that
is defined by $\phi_x(y)=(x,y)$ (or $\exp(2\pi i(x,y))$ in a multiplicative
notation).  This map is an isomorphism by general facts about lattices.

This is easily extended to cases that $G$ is a non-trivial cover of the
adjoint group.  If the adjoint group is $SO(4n)/\Z_2$, then
$G$ has fundamental group that is trivial or $\Z_2$, and either
way as in (1) above there is only one isomorphism of $\pi_1(G)$ with
its dual.  In other cases, the adjoint group has a cyclic fundamental
group $\Z_s$, for some $s$, and the fundamental group of $G$ consists
of elements of $\Z_s$ that are divisible by some divisor $t$ of
$s$.  Then by setting $\phi_x(y)=\exp(2\pi i(x,y)/t)$ we get
the desired natural isomorphism of $\pi_1(G)$ with its dual.

\subsec{Predictions For The Index}

We wish to describe the predictions for 
$\Tr \,(-1)^F$ and the more refined invariants $H^k(Q)$ that follow
from standard claims about the infrared behavior of this theory.
As before, we write $I(e,m)$ for the index as a function of $e$ and $m$.
We temporarily postpone discussion of the $H^k(Q)$ because of a subtlety
involving oblique confinement.

The first basic point is that if confinement holds, then there are
no zero energy states on a sufficiently big $\T^3$ for $e\not= 0$ and
$m=0$.  Hence
\eqn\jundo{I(e,0)=0, ~~{\rm for}~e\not= 0.}
By considering the case of simply-connected $G$, we have already deduced
that $I(0,0)=(-1)^r h$.
So \jundo\ completes the story for the index if $m=0$.

To understand what happens for $m\not= 0$, 
we begin as in section 2.2.  
We consider a path integral on $\T^4$ with supersymmetric
boundary conditions for fermions and with a $G$-bundle of 
characteristic class $\hat
m$.  We let $Z(\hat m)$ be the path integral for
this $G$-bundle.  We have   
\eqn\noggdo{Z(0)=(-1)^r h,}
since it equals $\Tr\,(-1)^F$ for the simply-connected cover of $G$.
We want to determine $Z(\hat m)$ for other $\hat m$'s.
If we regard the first three directions in $\T^4$
as the ``space'' directions, and the 
fourth direction as ``time,'' then it is natural to consider
the components $\hat
m_{ij}$ of $\hat
m$ with $i,j=1,\dots,3$ as ``magnetic''
and the $\hat
m_{i4}$ components as ``electric.''  We consider $\hat
m$ to
be ``purely magnetic'' if the electric components vanish, and
``purely electric'' if the magnetic components vanish.

Now, as in $2+1$ dimensions, we have the formula \ikl,
\eqn\nikl{I(e,0)={1\over |\Gamma|}\sum_{\gamma\in\Gamma}e^{-1}(\gamma)
Z(\hat
m_0(\gamma)).}
$\hat m_0(\gamma)$ is defined 
as in $2+1$ dimensions to have magnetic components
zero and electric components determined by $\gamma$.
Since $I(e,0)=0$ for $e\not= 0$, this formula implies that $Z(\hat
m_0(\gamma))$
is independent of $\gamma$.  Since $\hat 
m_0(\gamma)$ is an arbitrary purely
electric element of $H^2(\T^4,\pi_1(G))$, we find, using \noggdo\ to
fix the constant, that
\eqn\pikl{Z(\hat 
m)=(-1)^rh~{\rm if ~}\hat 
m{\rm ~is~purely~electric}.}

Now suppose that $\hat
m$ is purely magnetic.  The non-zero components of $\hat m$
are the 1-2, 1-3, and 2-3 components, but just as in section 2.3,
we can make an $SL(3,\Z)$ transformation to set $\hat
m_{12}=0$.  Then
after an $SL(4,\Z)$ transformation that exchanges the $3$ and $4$ directions,
$\hat
m$ becomes purely electric and we can use \pikl.
So we learn
\eqn\qpikl{Z(\hat 
m)=(-1)^rh~{\rm if ~}\hat 
m{\rm ~is~purely~magnetic}.}

The path integral $Z(\hat
m)$ for a purely magnetic $\hat
m$ has a natural
Hamiltonian interpretation.  
It is the index for states with magnetic flux $\hat m$.
This can be further decomposed according to the value of the electric
flux:
\eqn\ubikli{Z(\hat m
 )=\sum_eI(e,m).}
Hence 
\eqn\honflo{\sum_eI(e,m)=(-1)^rh}
for all $m$.
Just as in \ikli, the general
index can be written
\eqn\unikli{I(e,m)={1\over |\Gamma|}\sum_{\gamma\in\Gamma}e^{-1}(\gamma)
Z(\hat m
(m,\gamma)).}
(The definition of $\hat m(m,\gamma)$ is as before.  Writing 
$\T^4=\T^3\times \S^1$ with the last factor the ``time'' direction,
the magnetic part of $\hat m(m,\gamma)$ equals $m\in H^2(\T^3,\pi_1(G))$,
and the electric part is determined by $\gamma$.)

One might think that confinement would mean that $I(e,m)=0$ for all
nonzero $e$.  In fact, in $2+1$ dimensions, we proved this by
virtue of the assumption that $I(e,0)=0$ for nonzero $e$ plus $SL(3,\Z)$
symmetry.  However, in $3+1$ dimensions, confinement does not imply
that $I(e,m)$ always vanishes when $e$ is nonzero.

The physical reason for this has to do with 't Hooft's classification 
\thooft\
of  massive phases of gauge theories.  
\foot{As we have noted in a footnote above, this is not a complete
classification of conceivable
massive phases of gauge theories, but it describes the massive phases
that are believed to be relevant to the models under discussion here.}
According to this classification,
such phases are
described by condensation of a linear combination
of electric and magnetic charge.
If the condensed charge is electric, one has a Higgs phase; if the
condensed charge is magnetic, one has a conventional
confining phase.  If the
condensed charge is a mixture of electric and magnetic charge, 
one has a phase with ``oblique confinement.''
\foot{A slight elaboration of the classification, explained in  section 4
of \ref\donagi{R. Donagi and E. Witten, ``Supersymmetric Yang-Mills Systems
And Integrable Systems,'' hep-th/9510101, Nucl. Phys. {\bf B460} (1996) 299.},
 includes
``mixed phases,'' in which the gauge group is Higgsed to a subgroup that
is then confined or obliquely confined.  We will not recall the details
here as they are not relevant to the models studied in the present paper.}

In assuming that $I(e,0)=0$ for all non-zero $e$, we have incorporated
the idea that arbitrary external electric charge is confined.
Vanishing of $I(e,0)$ is expected whether the confinement is
ordinary or oblique confinement.  Once one turns on $m\not= 0$, however,
it is different; for $m\not= 0$, oblique confinement
might mean $I(e,m)\not = 0$ for certain nonzero $e$.  

To determine $I(e,m)$, we need to know, in the specific
$N=1$ supersymmetric gauge theories that we are studying, which
of the abstractly posssible massive phases are actually realized.
For this, we proceed as follows.  First of all, we recall from the derivation
of \jurry\ that the vacua of this theory are all of the form 
$v^s|\Omega\rangle$, where $v$ is a generator of the broken chiral
symmetry $\Z_{2h}$.  Moreover, the vacua are permuted in
an adiabatic increase of $\theta$ by $2\pi$.  
In other words, if one adiabatically increases $\theta$ by $2\pi$,
the state $v^s|\Omega\rangle$ is transformed to $v^{s-1}|\Omega\rangle$;
the vacua are transformed by $v^{-1}$.\foot{This standard
result is obtained as follows.
The one instanton amplitude is proportional to $e^{i\theta}\lambda^h$,
as there are $h$ zero modes of $\lambda$ in an instanton field.
Hence $\theta\to \theta+c$ is equivalent to $\lambda\to e^{ic/h}\lambda$.
So an adiabatic increase in $\theta$ by $2\pi$ gives back the same theory
with $\lambda\to e^{2\pi i/h}\lambda$, which is the action of $v^{-1}$.}

Now we recall 't Hooft's
intuitive explanation of oblique confinement: under a $2\pi$ increase
in $\theta$, a magnetic monopole may acquire electric charge
(as one sees explicitly in Higgs phases \ref\ewitten{E. Witten,
``Dyons Of Charge $e\theta/2\pi$,'' Phys. Lett. {\bf B86} (1979) 283.}).
In the present context, this means that under an adiabatic increase
of $\theta$ by $2\pi$, $e$ may not be invariant but will, in general,
change by an amount depending on $m$.  We will call this change in $e$ under
a $2\pi$ increase in $\theta$ the ``spectral flow'' and denote it by
$\Delta(m)$.
As we compute later,  $\Delta(m)$ is trivial for some groups,
and nontrivial for others.  But at any rate, we propose that
the appropriate physical prediction in this theory is
\eqn\plywood{I(e,m)=0{\rm~unless}~e~{\rm is~a~multiple~of}~\Delta(m).}
Thus, the phases are the ordinary confining phase and others obtained
from it by spectral flow.
Moreover, if $\Delta(m)$ is of order $c$ in the finite group $E$,
then we must have 
\eqn\nywood{I(e,m)=(-1)^r
{h\over c}~{\rm if}~e~{\rm is~a~multiple~of}~\Delta(m).}
In fact, since the different allowed $e$'s are obtained from each other
continuously by adiabatic increase in $\theta$, they must all have the same
index.

This obviously leaves us with the problem of computing $\Delta(m)$
for various $G$'s and $m$'s.
But first, we pause to extend our discussion from the index
to the more refined invariants, the $Q$ cohomology groups.

\bigskip\noindent{\it The Cohomology Groups}

We formulate the theory, in a Hamiltonian framework, on $\T^3$ with
a bundle of some fixed $m$.  We want to take $\T^3$ large and compare
to the bulk behavior on $\R^3$.  For this purpose, we should not project
onto a particular value of $e$, because the projection onto definite
$e$ is a global operation on $\T^3$.  Rather, we will sum over all $e$,
and construct the $Q$ cohomology in $\H_m=\oplus_e{\cal H}_{e,m}$.

We can reason just as we did at $m=0$.
Each vacuum of the infinite volume theory will, because of the mass gap,
contribute at most one supersymmetric state on $\T^3$.  As there are $h$
vacua, the total number of supersymmetric states is at most $h$.
On the other hand, the index formula \honflo\ says that the number of
supersymmetric states is at least $h$.  Hence the number is precisely
$h$, and they are in one-to-one correspondence with the vacua of
the infinite volume theory.

Those are, of course, permuted by the broken chiral symmetry.
Hence the reasoning that we used to get \ucases\ gives for all $m$
\eqn\nucases{h^k(Q;m)=\left\{\matrix{1,~{\rm for~even}~k+r\cr
         0,~{\rm for~odd}~k+r.\cr}\right.}
Here, of course, the $h^k(Q;m)$ are dimensions of the cohomology groups
for the action of $Q$ on the Hilbert space $\H_m$ with specified $m$.

Thus the picture at nonzero $m$ is similar to what it is at zero $m$.
There are $h$ supersymmetric states, each inherited from the bulk, and arising
at arbitrary even values of $k$.  If one wishes to further define
$Q$ cohomology groups in a sector in which both $e$ and $m$ are specified,
one runs into the following subtlety.  For groups $G$ such that $\Delta(m)=0$, 
one can define the $Q$ cohomology groups in each
$\H_{e,m}$, and these groups vanish for $e\not= 0$. 
It is different when there is oblique confinement.
The index $k$ labeling the cohomology groups is essentially
the eigenvalue of the generator $v$ of the chiral symmetry group $\Z_{2h}$.
To be more precise, a state $\psi\in H^k(Q)$ transforms under $v$
as $\psi\to e^{\pi ik/h}\psi$.  But $v$ is equivalent to 
 $\theta\to \theta-2\pi$, and this induces $e\to e - \Delta(m)$.  So one cannot
simultaneously measure both $k$ and $e$.  One way to proceed is to regard
$e$ as an element of $E/\Delta(m)$ (in other words, we consider $e$ and
$e'$ equivalent if $e-e'$ is a multiple of $\Delta(m)$),
and then  we have the prediction
\eqn\unucases{h^k(Q;0,m)=\left\{\matrix{1,\,\,{\rm for~ even}~ k+r\cr
0,\,\,{\rm for ~odd}~ k+r,\cr}\right. }
as well as $h^k(Q;e,m)=0$ if $e$ is a nonzero element of $E/\Delta(m)E$
(in other words, if $e$ is not divisible by $\Delta(m)$).

Now let us briefly discuss the physics of \nucases.
As 't Hooft showed, in a massive Higgs vacuum, a nonzero magnetic flux
on $\T^3$ produces a ``defect'' of strictly positive energy.  So a massive
Higgs vacuum would not contribute to $\Tr\,(-1)^F$ or the $H^k(Q)$.
But \nucases\ and our previous formulas for the index (both in $2+1$
and $3+1$ dimensions) show that for confining vacua, the magnetic flux
produces no such effect; it ``spreads out'' and produces no
contribution to the energy.

\bigskip\noindent{\it Action Of $SL(4,\Z)$}

In $2+1$ dimensions, we used $SL(3,\Z)$, plus the assumption
that $I(e,0)=0$ for all nonzero $e$, to prove that $I(e,m)=0$
for all nonzero $e$.  Here we would like to explain why one cannot
prove the analogous statement in $3+1$ dimensions by 
use of $SL(4,\Z)$.  

In $2+1$ dimensions, we proved in section 2.3
that by an $SL(3,\Z)$ transformation,
an arbitrary $m\in H^2(\T^3,\pi_1(G))$ can be made purely electric
(one can set $m_{12}=0$).  The key step was to argue that if $\pi_1(G)$
is a cyclic group $\Z_s$ for some $s$, then the only invariant of $m$
is the greatest common divisor, $\mu$, of the $m_{ij}$ and $s$.

\def\Pf{{\rm Pf}}
The difference in $3+1$ dimensions is that, even for $\pi_1(G)=\Z_s$,
it is not true that $\mu$ is the only invariant.  An element
$\hat m\in H^2(\T^4,\Z_s)$ has another invariant,
the ``Pfaffian''
\eqn\humbo{\Pf(\hat m)=\hat
m_{12}\hat m_{34}+\hat m_{13}\hat m_{42}+\hat m_{14}\hat m_{23}.}
It is well-defined modulo $\mu $ and related to the cup product as follows.
In general,
for $\hat m_1,\hat m_2\in H^2(\T^4,\Z_s)$, we define
\eqn\iko{(\hat m_1,\hat m_2)=\int_{\T^4}\hat m_1\cup \hat m_2.}
Then because $\T^4$ is spin, $(\hat m,\hat m)$ is divisible by 2 in a natural
fashion, and 
\eqn\doggo{\Pf(\hat m)={(\hat m,\hat m)\over 2}.}
It can be shown that $\mu$ and $\Pf(\hat m)$
classify $\hat m$ up to the action of $SL(4,\Z)$.\foot{The 
idea of the proof is to first reduce to $\mu=1$
by replacing $\hat m$ by $\hat m/\mu$ and $s$ by $s/\mu$.  Then one shows that
one can by an $SL(4,\Z)$ transformation set $\hat m_{12}=1$, after which
by an $SL(4,\Z)$ transformation one can set $\hat 
m_{ij}=0$ for $i=1,2$
and $j=3,4$ and (therefore) $\hat m_{34}=\Pf(\hat m)$.}

In view of \unikli, to get $I(e,m)=0$ for all $m$, one needs
$Z(\hat m(m,\gamma))$ to be independent of $\gamma$.  But 
$Z(\hat m(m,0))=(-1)^rh$
by virtue of \qpikl, so $Z(\hat m
  (m,\gamma))$ is independent of $\gamma$
 if and only if $Z(\hat m)$ is entirely independent of $\hat m$.

For $G$ and $     m $ such that the spectral flow  $\Delta(     
m)$ is nonzero, 
the instanton number is nonintegral,
as we will see, 
for certain $\hat m$,
and in particular cannot vanish.  Hence, the classical
action is strictly positive for such $\hat m$,
  and 
the partition function  vanishes for $g\to 0$ as $\exp(-{\rm const}/g^2)$.
Consequently, 
$Z(\hat 
m)=0$ for such $\hat 
m$; but $Z(0)=(-1)^rh$.  
So the fractionality of the instanton number is an obstruction to having
$Z(\hat m)$ independent of $\hat m$ and thus to having $I(e,m)$ vanish
for all nonzero $e$.
As we have already explained, and as we will see in more detail in section
3.4, the
correct  statement is that $I(e,m)=0$ unless $e$ is a multiple of $\Delta(m)$.
From our computations below, this assertion
is equivalent to the statement that $Z(\hat m
 )$ depends only on $\Pf(\hat m)$ and
not on $\mu$.  It would be interesting to try to verify this by direct
study of path integrals, but we will not do that in the present paper.
Section 4  of the paper is devoted to microscopic calculations
verifying the predictions that we have presented up to this point,
but these calculations will be done from a Hamiltonian point of view.

\subsec{Evaluation Of The Spectral Flow}

The purpose of the present section is to evaluate the spectral flow.
We consider a $G$-bundle over $\T^4$, with a given $\hat m
 \in H^2(\T^4,\pi_1
(G))$, and compute the deviation of the instanton number from being
an integer.  After doing this computation for $SU(n)/\Z_n$, we explain
why it determines the spectral flow.  Then
-- in a slightly lengthy case by case analysis -- we make the computation
for all simple Lie groups.

\bigskip\noindent{$SU(n)$}

We begin with $SU(n)$.
\def\Ad{{\rm ad}(V)}
 For an $SU(n)$ bundle $V$ on $\T^4$,
the instanton number or second Chern class is an integer $k$.
Let $\Ad$ be the  bundle derived from $V$ in the adjoint representation.
The ratio of the quadratic Casimir of the adjoint and fundamental
representations of $SU(n)$ is $2n$, so the instanton number of $\Ad$ is
$k'=2nk$.

Suppose now that there is a nonzero magnetic flux $\hat m$.  Then the gauge
group is really $SU(n)/\Z_n$, and the  $SU(n)$
bundle $V$ does not exist. The adjoint bundle $\Ad$ still exists,
and its instanton number is an integer, say $k'$.  The curvature integral
$(1/8\pi^2)\int \Tr F\wedge F$, which measures the instanton number for
an honest $SU(n)$ bundle,
assigns the value $k=k'/2n$ to $V$.  So the instanton number,
if measured in the usual units, is a rational number rather than an integer,
once there is a magnetic flux.

The value of the instanton number modulo 1 depends only on $\hat m$.
The reason for this is that any two $SU(n)/\Z_n$ bundles on $\T^4$ of the
same $\hat m$ can be obtained from one another by gluing in an ordinary
instanton, of integer instanton number.\foot{This statement can be proved
by obstruction theory, building up the bundle on the $r$-skeleton
of $\T^4$ for increasing $r$, and using the fact that the bundle
is determined up to the two-skeleton by $\hat m$, that $\pi_2(SU(n))=0$ so
nothing new happens on the three-skeleton,
and that $\pi_3(SU(n))=\Z$, which labels
 the extensions from the three-skeleton
over all of $\T^4$,  is associated with the instanton number.}
Thus to determine the value
of the instanton number, modulo $1$, for given $\hat m$, we need only
compute for a particular example.

\def\L{{\cal L}}
\def\O{{\cal O}}
\def\ad{{\rm ad}(V)}
Such an example can be constructed as follows. 
 Let ${\cal L}$ be a line bundle over $\T^4$
 whose first Chern class $c_1({\cal L})\in H^2(\T^4,\Z)$ reduces
 mod $n$ to $-\hat m$.
Now consider the ``$SU(n)$ bundle'' $V={\cal L}^{-1/n}\otimes(\L\oplus \O
\oplus \O\oplus \dots \oplus \O)$, with $n-1$ trivial line bundles ${\cal O}$.
Because of the fractional exponents, $V$ does not really exist
as an $SU(n)$ bundle.  However, the associated adjoint bundle $\ad$
does exist as an $SU(n)/\Z_n$ bundle.\foot{Concretely, $\ad$ is the
direct sum of $n-1$ copies of $\L$, $n-1$ copies of $\L^{-1}$,
and $(n-1)^2$ copies of ${\cal O}$.}  Its magnetic flux equals $\hat m$.
(The basic reason for this is that 
the nonzero value of $-c_1({\cal L})/n$ mod 1 or equivalently
of $\hat m= -c_1({\cal L})$ mod $n$ is the obstruction to defining ${\cal L}^{-1/n}$
as a line bundle; this is the same as the obstruction to defining $V$ as
a vector bundle since ${\cal L}^{-1/n}$ is the only problematic factor
in the definition of $V$.)

 Note that \eqn\kipl{\int_{\T^4}
c_1(\L)\cup c_1(\L)=2\hat m_{12}\hat m_{34}=2\,\Pf(\hat m).}
The instanton number of $V$
(which as noted above is $1/2n$ times the instanton number of $\ad$)
can be correctly computed, despite the fractional exponents in the
definition of $V$,
using the fact that for any $SU(n)$ bundle $W=\oplus_{i=1}^n\L_i$,
the instanton number is 
\eqn\hcn{\int_{\T^4}c_2(W)=\sum_{i<j}\int_{\T^4} c_1(\L_i)\cup c_1(\L_j).}
Using \hcn\ and \kipl, a small computation shows that the instanton
number of $V$ is
\eqn\jcn{k=-\Pf(\hat m)\left(1-{1\over n}\right).}
We will denote the deviation of the instanton number from being
an integer as $\Delta'(\hat m)$.
So
\eqn\hch{\Delta'(\hat m)={\Pf(\hat m)\over n}\,\,{\rm modulo}~1}
for $SU(n)$.

Our computations below for other groups
will always involve reducing to an $SU(n)$ subgroup
for some $n$.  It will help to introduce the following notation.
For any $n\geq 2$ and $\hat m\in H^2(\T^4,\Z_n)$, we let $V_n(\hat m)$ be
an ``$SU(n)$ bundle'' whose existence is obstructed by the magnetic
flux $\hat m$.  This means, to be more precise, that ${\rm ad}(V_n(\hat m))$
is an $SU(n)/\Z_n$ bundle of magnetic flux $\hat m$.

\bigskip\noindent{\it Relation To The Spectral Flow}

Before computing the deviation $\Delta'(\hat m)$
 from integral instanton number for other
groups, we will explain why this deviation determines the spectral flow
$\Delta( m)$.

Consider first a $G$-bundle $X$ over $\T^3$ with some magnetic
flux $m\in H^2(\T^3,\pi_1(G))$ and some connection $A$.  
The connection has a Chern-Simons invariant
\eqn\icco{CS(A)={1\over 8\pi^2}\int_{\T^3}\Tr\left( 
A\wedge dA+{2\over 3}A\wedge A\wedge A
\right).}
$CS(A)$ is invariant under gauge transformations that are continuously
connected to the identity but not under all gauge transformations.
It has been normalized so that it
is shifted by 1 under the gauge transformation $T$ associated with the
generator of $\pi_3(\hat G)=\Z$.\foot{This definition of $CS(A)$
assumes  that the bundle is trivial and so $A$ can be regarded globally
as a Lie algebra valued one-form.  More generally, we let $X$ be a 
four-manifold of boundary $\T^3$ over which 
the bundle and connection extend, and 
we define $CS(A)=(1/8\pi^2)\int_X\Tr F\wedge F$.}

$CS(A)$ is not a topological invariant.  But it has the following
very important property. 
If $A$ is flat, that is if $F=dA+A \wedge A$ is zero, then $CS(A)$
is invariant under continuous
deformations of $A$ that preserves the flatness.
For under any first order deformation $A\to A+\delta A$, the variation
of $CS(A)$ is a multiple of $\int \Tr\,\delta A\wedge F$, which vanishes
for $F=0$.

Let $\tilde\Gamma$ be the group of components of the group of
gauge transformations of $X$.
$\pi_3(G)$ gives some information about $\tilde \Gamma$, but
if the gauge group is not simply-connected,
$\pi_1(G)$ also enters in describing $\tilde\Gamma$.  As we will
now explain, the two mix in an interesting way. 
Let $\gamma$ be a gauge transformation whose restriction to loops
in $\T^3$ determines an element, which we will call $\bar \gamma$,
of $\Gamma=\Hom(\pi_1(\T^3),\pi_1(G))$.
To determine how $CS(A)$ transforms under $\gamma$, we proceed as follows.
Rather as in the example sketched
in figure 1, we can use the gauge transformation $\gamma$ as gluing
data in the  $\S^1$ or ``time''
direction to extend the $G$-bundle $X\to \T^3$ to a $G$-bundle
$X'\to \T^4=\T^3\times \S^1$.  This bundle has a characteristic class in $H^2(\T^4,\pi_1(G))$
that we write
as $\hat m(m,\gamma)$; its magnetic components coincide with $m$, and the
electric components are determined by $\gamma$.
Moreover, the change $\Delta_\gamma(CS(A))$
in $CS(A)$ under $\gamma$ is simply
the instanton number of the bundle $X'$.  This is $\Delta'(\hat m)$ modulo 1:\foot{The reason that we only
get a nice formula for $\Delta_\gamma(CS(A))$ mod 1 is that to
determine $\Delta_\gamma$ precisely, we would need a precise description
of the gauge transformation $\gamma$, and not just its image $\bar\gamma$
in $\Gamma={\rm Hom}(\pi_1(\T^2),\pi_1(G))$.  Without changing
$\bar\gamma$, we could substitute $\gamma\to\gamma T^k$ for any
integer $k$; this would add $k$ to $\Delta_\gamma(CS(A))$.}
\eqn\ucdu{\Delta_\gamma(CS(A))=\Delta'(\hat m)~{\rm modulo}~1.}

Now suppose that $\gamma$ is such that $\bar \gamma$ is of order $s$.
Thus $\gamma^s$ determines a trivial homomorphism of $\pi_1(\T^3)$ to
$\pi_1(G)$.  This does not mean that $\gamma^s$ is homotopic to the identity
gauge transformation; it only means that up to homotopy
\eqn\nixy{\gamma^s=T^r}
for some $r$.  We can determine $r$ modulo $s$ by noting that $\gamma^s$
shifts $CS(A)$ by $s \Delta_\gamma(CS(A))=s \Delta'(\hat m)$ mod $s$.
Since $T$ shifts $CS(A)$ by 1, we have
\eqn\umixy{r=s\Delta'(\hat m)\,\, {\rm modulo}~s.}
The undetermined multiple of $s$ in \umixy\ is inessential, 
as $r$ can be shifted by
a multiple of $s$ by $\gamma\to \gamma T^k$, as in the footnote.

If $r$ is not always zero modulo $s$, 
we have the following mathematical situation.
The group $\tilde \Gamma$  of all homotopy classes of gauge 
transformations of the bundle $X\to \T^3$ maps
to $\Gamma$ by mapping a gauge transformation $\gamma$ to its image
$\bar \gamma\in \Gamma$.  It contains a subgroup $\Z=\pi_3( G)$ of gauge
transformations that map trivially to $\Gamma$.  However, if $\Delta'(\hat m)$
is not zero for some $\gamma$, then
$\tilde \Gamma$ is not a product but a nontrivial extension:
\eqn\ilpu{0\to \Z\to\tilde \Gamma\to \Gamma\to 0.}
Indeed, since $T$ is a generator of $\Z$, the relation
$\gamma^s=T^r$ has no solution in $\Z\times\Gamma$ if $r$ is not a multiple
of $s$, so having a $\gamma\in\tilde\Gamma$ that obeys this relation
means that $\tilde \Gamma$ is a nontrivial extension rather than a product.

Now, we are almost ready to compute the spectral flow.  In this situation,
we must be more careful about defining the electric flux $e$.
So far (keeping to $\theta=0$) we have regarded the electric flux
as a character of the group $\Gamma$.  If we want to generalize this
to arbitrary $\theta$, we should instead regard $e$ as a character of
the full group $\tilde \Gamma$.  However, to define the theory at given
$\theta$, we want to allow only those characters $e$ which on the subgroup
$\Z$ act as follows: $T$ is mapped  to $e^{i\theta}$.

To compute the spectral flow, we fix a character $e$ of $\tilde \Gamma$
for fixed $\theta$, and follow it continuously under $\theta\to\theta+2\pi$.
Let us take  an arbitrary $\gamma\in \tilde \Gamma$ and follow
$e(\gamma)$ while $\theta$ is increased.  We have from the above
$\gamma^s=T^r=T^{s\Delta'}$, so $e(\gamma)^s=e(\gamma^s)=e(T^{s\Delta'})=
e(T)^{s\Delta'}
=e^{i\theta s\Delta'}$.  Taking the $s^{th}$ root of this, we learn
that 
\eqn\limmy{e(\gamma)= C e^{i\theta\Delta'},}
where $C$ is a root of unity that is
independent of $\theta$.  Hence under $\theta\to\theta+2\pi$,
we have
\eqn\rimmy{e(\gamma)\to e(\gamma) e^{2\pi i\Delta'}.}
Switching to an additive notation, this means that under $\theta\to \theta
+2\pi$, $e$ is shifted by
\eqn\polo{e\to e+\Delta(m)}
where $\Delta(m)$ is defined as follows: 
$\Delta(m)$ maps $\gamma\in \tilde \Gamma$ to $\Delta'(\hat m(m,\gamma))$ or
more briefly to $\Delta'(m,\gamma)$.

More informally, we might say that $\Delta=\Delta'$; the deviation $\Delta'$
from integral instanton number determines the spectral flow $\Delta$.
In the remainder of this section, we compute $\Delta'$ and hence $\Delta$
for the remaining simple Lie groups, starting with the symplectic group.

\bigskip\noindent{$Sp(n)$}

  The center of $Sp(n)$ is $\Z_2$, generated
by the element that acts as $-1$ on the fundamental $2n$-dimensional
representation of $Sp(n)$.  The adjoint group is therefore $G=Sp(n)/\Z_2$,
with $\pi_1(G)=\Z_2$.

A simple example of an $Sp(n)/\Z_2$ bundle with given $\hat m$ can be constructed
as follows.  For $n=1$, $Sp(1)=SU(2)$, and we take the $SU(2)/\Z_2$ bundle
${\rm ad}(V_2(\hat m))$ that was defined in the discussion of $SU(n)$.
  Now, for general $n$,
$Sp(n)$ contains the subgroup $SU(2)^n$ generated by the ``diagonal
quaternions.''  Consider the ``$Sp(n)$ bundle'' $W=\oplus_{i=1}^nV^{(i)}_2(\hat
m)$,
with $V_2^{(i)}(\hat m)$ a copy of $V_2(\hat m)$ 
for the $i^{th}$ factor in $SU(2)^n\subset Sp(n)$.
Once again, it is only the associated adjoint bundle ${\rm ad}(W)$ that
really exists globally.  
\foot{
In fact, ${\rm ad}(W)=\oplus_i {\rm ad}(V_2^{(i)}(\hat m))\oplus_{i<j}V_2^{(i)}(\hat m)
\otimes V_2^{(j)}(\hat m)$,
which is isomorphic to the sum of $n$ copies of ${\rm ad}(V_2(\hat m))$
and  $n(n-1)/2$ copies of ${\cal O}$.  This is worked out by looking
at the decomposition of the adjoint representation of $Sp(n)$ under
$SU(2)^n$.}  The characteristic
class that restricts the lifting of  ${\rm ad}(W)$ to $W$ is $\hat m$, since
this is the characteristic class that obstructs lifting the 
${\rm ad}(V_2^{(i)}(\hat m))$ to $SU(2)$ bundles
$V_2^{(i)}(\hat m)$; defining $W$ is of course equivalent
to defining the summands $V_2^{(i)}(\hat m)$.

Using \hch, the instanton number of $W$ receives
a contribution $\Pf(\hat m)/2$ for each $SU(2)$ factor, so for $Sp(n)$ we get
\eqn\nch{\Delta'(\hat m)={n\over 2}\Pf(\hat m)~{\rm modulo}~1.}
So oblique confinement occurs
in $Sp(n)$ for $n$ odd but not for $n$ even.

\bigskip\noindent{$E_6$}

We consider next the exceptional group $E_6$.  The center is $\Z_3$,
and the adjoint group is $G=E_6/\Z_3$.

$E_6$ contains a subgroup locally isomorphic to
$SU(3)\times SU(3)\times SU(3)$.  
An $E_6/\Z_3$ bundle $X$ with given $\hat m$ can be constructed from the
bundle that is respectively $V_3(0)$, $V_3(-\hat m)$, and $V_3(\hat m)$ for the
three $SU(3)$'s.
First of all, to see that this does give an $E_6/\Z_3$ bundle,
we look at the decomposition of the adjoint representation of $E_6$
under $SU(3)\times SU(3)\times SU(3)$:
\eqn\plimpo{{\bf 78}=({\bf 8},{\bf 1},{\bf 1})\oplus ({\bf 1},{\bf 8},{\bf 1})
\oplus ({\bf 1},{\bf 1},{\bf 8})\oplus ({\bf 3},{\bf 3},{\bf 3})
\oplus (\overline {\bf 3},\overline{\bf 3},\overline {\bf 3}).}
With the given $SU(3)\times SU(3)\times SU(3)$ 
bundle, we see that our adjoint bundle
is 
\eqn\kilpp{{\rm ad}(V_3(0))\oplus {\rm ad}(V_3(-\hat m))\oplus 
{\rm ad}(V_3(\hat m))
\oplus V_3(0)\otimes V_3(-\hat m)\otimes V_3(\hat m)\oplus \bar {V_3(0)}\oplus
\bar{V_3(-\hat m)}\oplus \bar{V_3(\hat m)}.}  
(For a bundle $W$, $\bar W$ is the
complex conjugate bundle; $\bar W$ has opposite magnetic flux to $W$.)  
This is a well-defined $E_6/\Z_3$ bundle,
because the total magnetic flux vanishes for each summand.
To show that the magnetic flux of $X$ is $\hat m$, we note 
that the ${\bf 27}$ of $E_6$ decomposes
under $SU(3)^3$ as 
\eqn\hipo{({\bf 3},\overline{\bf 3},{\bf 1})\oplus ({\bf 1},{\bf 3},
\overline {\bf 3})\oplus(\overline{\bf 3},{\bf 1},{\bf 3}).}
Hence, a hypothetical lift of $X$ to a bundle whose fiber transforms
in the ${\bf 27}$ of $E_6$ would be 
\eqn\kicv{X'=V_3(0)\otimes \bar {V_3(-\hat m)}\oplus
V_3(-\hat m)\otimes \bar{V_3(\hat m)}\oplus V_3(\hat m)\otimes \bar{V_3(0)}.}
Since $\bar{V_3(-\hat m)}=V_3(\hat m)$, and $3\hat m=0$ 
(as the fundamental group is
$\Z_3$), the ``total magnetic flux'' of each summand is $\hat m$.  Hence,
using any of the constructions of nontrivial bundles given at the end of
section 2.1, the obstruction to lifting $X$ to an $E_6$ bundle $X'$
is the same as the obstruction to lifting ${\rm ad}(V_3(\hat m))$ to an $SU(3)$
bundle $V_3(\hat m)$.  The characteristic class of $X$ is therefore $\hat m$.
When we consider other groups below, we will not describe the sort of arguments
presented in the present paragraph in such detail.

The instanton number of the $E_6/\Z_3$ bundle $X$, 
according to \hch, receives
contributions $0,$ $\Pf(-\hat m)/3=\Pf(\hat m)/3$, 
and $\Pf(\hat m)/3$ from the three $SU(3)$'s, so
\eqn\dutty{\Delta'(\hat m)={2\over 3}{\Pf(\hat m)}~{\rm modulo } ~1}
for $E_6/\Z_3$.
$E_6$ will therefore exhibit oblique confinement.

\bigskip\noindent{$E_7$}

Next we consider the other exceptional group with a nontrivial center,
which is $E_7$, with $\pi_1(E_7)=\Z_2$.  The adjoint group is hence
$G=E_7/\Z_2$.

$E_7$ contains a subgroup that is locally $SU(4)\times SU(4)\times SU(2)$.
The ${\bf 56}$ transforms as
\eqn\dicco{({\bf 4},\overline {\bf 4},{\bf 1})\oplus (\overline {\bf 4},
{\bf 4},{\bf 1})\oplus ({\bf 6},{\bf 1},{\bf 2})\oplus ({\bf 1},{\bf 6},
{\bf 2}).}
A convenient $E_7/\Z_2$ bundle of given $\hat m$ can be constructed by taking
the ``$SU(4)\times SU(4)\times SU(2)$ bundle'' $V_4(2\hat m)\otimes V_4(0)
\otimes V_2(\hat m)$.  (Note that as $\hat m$ is defined modulo 2, $2\hat m$ is defined
modulo 4, so $V_4(2\hat m)$ is defined.)  
That this $E_7/\Z_2$ bundle is well-defined
and has magnetic
flux $\hat m$ may be argued along lines sketched above for $E_6$.
Using \hch, we learn that the three factors contribute 
$\Pf(\hat m)$, 0, and $\Pf(\hat m)/2$
to the instanton number, so for $E_7/\Z_2$,
\eqn\kicco{\Delta'(\hat m)={\Pf(\hat m)\over 2}~{\rm mod}~1.}
Hence there will be oblique confinement for $E_7$.

\bigskip\noindent{$Spin(2n+1)$}

We consider next the case of $Spin(2n+1)$. The center is $\Z_2$,
and the adjoint group is $G=SO(2n+1)$.

To find an unliftable $SO(2n+1)$ bundle with given $\hat m$, we proceed
as follows.  Starting with a subgroup 
\eqn\hy{SO(3)\times SO(2n-2)\subset
SO(2n+1)} and observing that $SO(3)=SU(2)/\Z_2$, we take the $SU(2)$
bundle $V_2(\hat m)$ times a trivial $SO(2n-2)$ bundle.  According to \hch,
the instanton number of $V_2(\hat m)$ as an $SO(3)$ instanton
is $\Pf(\hat m)/2$.  (In other words, its instanton number is $\Pf(\hat m)/2$ 
times the smallest instanton number of an $SO(3)$ instanton on $\S^4$,
which is conventionally called 1.)
 However, an $SO(3)$ instanton
of instanton number 1 has instanton number 2 when embedded in 
$SO(2n+1),~n\geq 2$
via the embedding \hy.\foot{An 
$SO(2n+1)$ field of instanton number 1 on $\S^4$ is made via the embedding
$ SU(2)\times SU(2)
=SO(4)\subset SO(2n+1)$ by placing a field of instanton number 1 in 
the first $SU(2)$.  Placing instanton number 1 in the $SO(3)$ in \hy\
amounts to putting instanton number 1 in each of the two $SU(2)$'s,
which gives instanton number 2 in $SO(2n+1)$.  Hence, the ``quantum''
of instanton charge on $\S^4$ is half as big for $SO(n)$ with $n>3$
as for $SO(3)$.}  Hence, the $SO(2n+1)$
instanton number is $\Pf(\hat m)$, and vanishes modulo 1.  So
\eqn\snin{\Delta'(\hat m)=0~{\rm modulo}~1}
for $Spin(2n+1)$, and $SO(2n+1)$ has no oblique confinement.
In other words, if $SO(2n+1)$ $N=1$ super Yang-Mills theory is probed
by external 't Hooft and Wilson loops in the spinor representation of
the gauge group, one will detect ordinary rather than oblique confinement.

The unliftable bundle that we constructed above makes sense for
$SO(2n)$ as well as for $SO(2n+1)$.
However, in contrast to $Spin(2n+1)$, whose center is $\Z_2$,
$Spin(2n)$ has center $\Z_2\times \Z_2$ or $\Z_4$ for even or odd $n$,
so to give a complete answer for $Spin(2n)$ requires further
consideration that we present below.
But if one considers a $Spin(2n)$ theory with dynamical charges
in the vector representation of $SO(2n)$, then the center is effectively
reduced to $\Z_2$.
We can accomplish this by taking the minimal $N=1$ super Yang-Mills theory
with gauge group $SO(2n)$ and adding chiral superfields in the vector
representation with a large (supersymmetric) bare mass.
The above computation shows that the relevant spectral flow vanishes,
so this theory should have  ordinary rather than
oblique confinement in all the vacua.
 
As has been pointed out by N. Seiberg, this assertion
can be verified in another way using
dynamical results in  \ref\seint{K. Intriligator and N. Seiberg, ``Duality,
Monopoles, Dyons, Confinement And Oblique Confinement In
Supersymmetric $SO(N(C))$ Gauge Theories,'' Nucl. Phys. {\bf B444} (1995) 125,
hep-th/9503179.}.  Consider an $N=1$ supersymmetric
theory with gauge group $SO(k)$ and $k-2$ chiral superfields $Q_i$
in the vector representation; let $M_{ij}$
be the ``meson fields'' $Q_i\cdot Q_j$ and set $U= \det M$. The $U$ plane
parametrizes a Coulomb phase.  (This statement depends on the fact that
the number of chiral multiplets is exactly $k-2$.)
We assume $k\geq 5$ to avoid some special
phenomena at small $k$, but it will not matter whether $k$ is odd or even.
There are  two points on the $U$ plane
with massless monopoles or dyons.  At one point {\it (A)}
there are massless ``dyons'' $q^i_\pm, \,i=1,\dots,k$ with a coupling 
$M_{ij}q^i_+q^j_-$, and at a second locus {\it (B)} there is a single pair
of massless ``monopoles'' $S_\pm$ with a coupling $S_+S_-(\det M-1)$.  
The reason that we have
put the words ``monopoles'' and ``dyon'' in quotes in the last sentence
is that it has not been completely clear which of these particles
are monopoles and which are dyons, but the analysis in \seint\ shows
that the nonperturbative massless particles in the {\it (A)} and {\it (B)}
vacua are of opposite types.  To perturb down to the pure
$SO(k)$ theory, one adds a generic quark mass term, corresponding to a 
perturbation $\Tr \,mM$ of the superpotential. 
For large $m$, the theory reduces to the pure $SO(k)$ theory, but
on the other hand, the value of $m$ does not affect the vacuum structure.
Upon turning on $m$, most of the $U$ plane including
the  {\it (A)} vacuum disappears; the {\it (B)} vacuum
splits into $k-2$ massive vacua, which (as $k-2$ equals the dual Coxeter
number  of $SO(k)$ for $k\geq 5$) is the correct
number to saturate $\Tr\,(-1)^F$ for $SO(k)$.  
All vacua, as they arise from {\it (B)}, are of the same type,
showing (in agreement with the direct computation above)
that the minimal $N=1$ theory with $SO(k)$ gauge group 
must have no spectral flow as long as we only consider bundles that
are well-defined in the vector representation.  Moreover, in view
of our microscopic knowledge about this $SO(k)$ theory, this shows that
the massless excitations at the point {\it (B)} really are monopoles,
corresponding to ordinary confinement.

\bigskip\noindent{\it $Spin(4n+2)$}

We will now carry out more complete analyses for $Spin(4n+2)$ and $Spin(4n)$,
including the bundles that do not lift to the vector representation of
$SO(4n+2)$ or $SO(4n)$.  First we consider $Spin(4n+2)$.

The center of $Spin(4n+2)$ is $\Z_4$; the adjoint group is $G=Spin(4n+2)/\Z_4$.

For $n=1$, we have $Spin(6)=SU(4)$.  For an $SU(4)/\Z_4$
bundle of magnetic flux $\hat m$,
we can use our friend $V_4(\hat m)$ (or more precisely its adjoint bundle).
The instanton number of $V_4(\hat m)$ is $\Pf(\hat m)/4$.

To study $Spin(4n+2)$, we use the subgroup
\eqn\rly{Spin(6)\times Spin(4)\times
\dots \times Spin(4),} with $n-1$ factors of $Spin(4)$.
A $Spin(4n+2)/\Z_4$ bundle of magnetic flux $\hat m$ can be constructed
using the $Spin(6)$ bundle $V_4(\hat m)$, and the bundles 
$V_2(2\hat m)\otimes V_2(0)$
for each of the $Spin(4)=SU(2)\times SU(2)$ 
factors in \rly.  (As $2\hat m$ is of order 2, $V_2(2\hat m)$ is defined.)
This is chosen  so that the two spinor representations of $Spin(4n+2)$
are obstructed by $\hat m$ and $-\hat m$, respectively, and the vector
representation by $2\hat m$.
The instanton number of
$V_2(2\hat m)\otimes V_2(0)$ is integral, and that of $V_4(\hat m)$ is
$\Pf(\hat m)/4$, so altogether we have
\eqn\hbby{\Delta'(\hat m)={\Pf(\hat m)\over 4}~{\rm modulo }~1}
for $SO(4n+2)$.
So $Spin(4n+2)$ has oblique confinement.   

Let us now reconcile this with the assertion that if massive
particles in the vector representation are added, one does not
see oblique confinement.  A formal way to say things is that
in a theory that contains
particles in the vector representation of $SO(4n+2)$, we must take $\hat m$
to be divisible by 2, and then $\Delta'(\hat m)$ vanishes and we do not
observe oblique confinement.

A more physical (and essentially standard) explanation is as follows.
First we recall in detail the meaning of oblique confinement.
Formulating the theory on a spatial manifold
$\R^3$, we let, for $C$ a loop in $\R^3$,
$W(C)$ denote a Wilson loop operator
for an external charge propagating around $C$ in the positive chirality
spinor representation of $Spin(4n+2)$.
We also let $H(C)$ be an 't Hooft loop constructed with a gauge transformation
on $\R^3-C$ whose monodromy around $C$ is a generator of the 
center $\Z_4$ 
of $Spin(4n+2)$.  If $C$ and $C'$ are loops in $\R^3$ of linking
number 1, one has the 't Hooft algebra
\eqn\polo{W(C)H(C')=\exp(2\pi i/4)H(C')W(C).}
In a confining vacuum, there is an integer $s$ such that
the  operator $H(C)W^s(C)$ has no area law (and $H^rW^t$ does unless
$t=sr$ mod 4); 
oblique confinement is the case that $s\not= 0$.
If we add massive charges in the vector representation, then $H(C)$ no
longer makes sense as an observable, but $H(C)^2$ does.  In addition,
we should consider $W(C)^2$ to be trivial since it can be screened by
a charge in the vector representation.  The basic observables
are thus $H(C)^2$ and $W(C)$, and they obey an 't Hooft algebra
appropriate for a group $SO(4n+2)$ with fundamental group $\Z_2$:
\eqn\inxon{H(C)^2W(C')=-W(C')H(C)^2.}
The vacuum that formerly was described as having no area law
for $H(C)W^s(C)$ must now (since only even powers of $H(C)$ are defined)
be described as having no area law for $H(C)^2W^{2s}(C)$; since
$W^{2s}(C)$ is trivial, this is equivalent to saying that $H(C)^2$ has
no area law.  The information about $s$ is lost.

\bigskip\noindent{\it $Spin(4n)$}

The last example is $Spin(4n)$, with center $\Z_2\times \Z_2$.
The adjoint group is $G=Spin(4n)/\Z_2\times \Z_2$.

We select generators $a_1$, $a_2$ of the center of $Spin(4n)$
 defined as follows.
$a_1$ is $+1$ on the positive chirality spinor representation, $-1$
on the negative chirality spinor representation, and $-1$ on the vector
representation; $a_2$ acts on those three representations as $-1,+1,-1$.
Equivalently, we can regard $a_1$ and $a_2$ as generators of $\pi_1(G)$.
Since $H^2(\T^4,\pi_1(G))=H^2(\T^4,\Z_2)\otimes_{\Z_2}\pi_1(G)$,
we can write an arbitrary $\hat m\in H^2(\T^4,\pi_1(G))$  
as $\hat m=\hat m_1a_1+\hat m_2a_2$, with  $\hat m_1,\hat m_2\in 
H^2(\T^4,\Z_2)$.

The first example of $Spin(4n)$ is $Spin(4)=SU(2)\times SU(2)$.
A $Spin(4)/\Z_2\times \Z_2$ bundle of general $\hat m$ is derived from
$V_2(\hat m_1)\otimes V_2(\hat m_2)$.
The instanton number is $\half(\Pf(\hat m_1)+\Pf(\hat m_2)).$

For general $n$, we look at the subgroup 
\eqn\ilopo{Spin(4)\times Spin(4)\times \dots \times Spin(4)\subset Spin(4n),}
with $n$ factors.  If $n$ is odd, we use $V_2(\hat m_1)\otimes V_2(\hat m_2)$
in each factor, and get instanton number $(n/2)(\Pf(\hat m_1)+\Pf(\hat m_2))$,
 so 
\eqn\kipono{
 \Delta'(\hat m)=
{\Pf(\hat m_1)+\Pf(\hat m_2)\over 2}~{\rm modulo}~1}
for $Spin(8k+4)$.  
The choice of bundle has been made to ensure that the
total magnetic flux obstructing the existence of the positive or negative
chirality spin bundles is $\hat m_1$ or $\hat m_2$, respectively.
If $n$ is even, to achieve the same end,
we build the bundle using
$V_2(\hat m_1+\hat m_2)\otimes V_2(0)$ in the first
factor and $V_2(\hat m_1)\otimes V_2(\hat m_2)$ in the others.  
Then we get
\eqn\ilpoo{\Delta'(\hat m)={(\hat m_1, \hat m_2) \over 2}~{\rm modulo } ~1}
for $Spin(8k)$.  
(We have used the fact that in view of \doggo,
$(\hat m_1,\hat m_2)=\Pf(\hat m_1+\hat m_2)
-\Pf(\hat m_1)-\Pf(\hat m_2)$.)
  Thus, $Spin(8k)$ and $Spin(8k+4)$ both
have oblique confinement, with somewhat different details.

In a theory in which there are dynamical charges
in the vector representation of $Spin(4n)$, we must set $\hat m_1=\hat m_2$,
and then $\Delta'(\hat m)=0$ so we do not observe oblique confinement.
(For $Spin(8k)$, this depends on the fact that $(\hat m,\hat m)$ is always
even.)  
One can give a more physical explanation of this as we did for $Spin(4n+2)$.

\subsec{Outer Automorphisms}

We now consider the case
that  $G$ admits a group $C$ of outer automorphisms.
 In practice, for $G$ a simple Lie     group,
$C$ will be $\Z_2$, except in the case of $G=Spin(8)$, where $C$
can be the triality  group (the group of permutations of three elements)
or a subgroup.  For example, concretely, for $G=SU(n)$, 
$C=\Z_2$ is generated by the
operation of complex conjugation or ``charge conjugation'' that exchanges
the representations ${\bf n}$ and ${\bf {\bar n}}$ of $SU(n)$.

 Such a group of outer automorphisms of the gauge group  is
realized in the quantum field theory as a global symmetry group,
and one would like to know if $C$ is spontaneously broken.
  We 
expect that it is not, since the usual logic of confinement does
not distinguish representations of $G$ that differ by the action of $c$,
so we do not expect $c$ to be spontaneously broken.  For example,
for $G=SU(n)$, we do not expect spontaneous breaking of charge conjugation
symmetry.  

There are two ways that we can formulate a criterion for $C$ to
be unbroken in terms of generalizations of $\Tr\,(-1)^F$.  
First of all, because $C$ commutes with supersymmetry, we can pick
any element $c\in C$ and define $I(c)=\Tr\, c(-1)^F$.  This trace
can be evaluated just in the space of zero energy states, as the 
states of nonzero energy cancel in bose-fermi pairs.  
If then the $h$ vacua of the theory are all $C$-invariant in infinite
volume, then we expect
\eqn\unijx{I(c)=(-1)^rh,}
just as at $c=0$.  We can generalize this to include the magnetic
and electric flux $m$ and $e$. The only subtlety is that they must
be $c$-invariant in order for $c$ to act in the Hilbert space with specified
$m$ and $e$.  This condition can be very restrictive;
for example, for $G=SU(n)/\Z_n$, the charge conjugation automorphism
$c$ acts by $m\to -m$, $e\to -e$.  At any rate, for such $m$ and $e$,
we expect $I(e,m;c)=\Tr_{{\cal H}_{e,m}}c(-1)^F$  to be independent
of $c$ as it can be computed from $C$-invariant vacuum states.

The second approach to formulating a criterion for $C$ to be unbroken
is to consider the gauge theory formulated on a bundle that is twisted
by some element $c\in C$.
We can combine $C$ and $G$ to a disconnected gauge group $G'$
that fits into an exact sequence
\eqn\uttu{0\to G\to G'\to C \to 0.}
We think of $G'$ as the gauge group and formulate the theory
using a non-trivial $G'$ bundle over $\T^3$.  This is analogous
to the way the magnetic flux was introduced when $\pi_1(G)\not=0$,
and just as in that case, in the quantization we want to require the
quantum states
to be invariant only under a restricted group of gauge transformations
-- the gauge transformations by elements of $G$.

This procedure is easy to carry out.
If $G$ is simply-connected, a $G'$ bundle on $\T^3$
can be classified very simply.  The holonomy around a circle
$S\in \T^3$ is an element of $G'$ whose image in $C$ depends only
on the class of $S$ in $\pi_1(\T^3)$. This gives a homomorphism
$\phi:\pi_1(\T^3)\to C$, which completely classifies the $G$-bundles.
Since $\pi_1(\T^3)$ is commutative, the image of $\phi$ is a commutative
and hence cyclic subgroup of $C$ in all cases (even $G=Spin(8)$).
For a suitable decomposition $\T^3=\S^1\times \T^2$, the bundle
can be described as follows: there is some $c\in C$, such that
in going around $\S^1$, the   fields are conjugated by $c$.  We let
$X_c$ denote the bundle constructed in this way, and we write
$I_c$ for the index $\Tr\,(-1)^F$ for the supersymmetric gauge theory
quantized using the bundle $I_c$.

We evaluate $I_c$ as usual by taking a large metric on $\T^3$ and saturating
the trace by zero energy vacuum states.  The only subtlety is that,
when we quantize on $X_c$, a vacuum state of the infinite volume theory
that is not $c$-invariant will not contribute, since in trying to glue
in such a state on $\T^3$, one will need a domain wall (because of the twist
by $c$) with a large cost in energy.  If, however, all vacuum states
of the infinite volume theory are $C$-invariant and massive,
they should all contribute in the quantization on $X_c$, and
so 
 we expect
\eqn\inkon{I_c=(-1)^rh~{\rm for~all}~c.}

We can extend this to the case that $G$ is not simply-connected, so
that the bundles are classified by a magnetic flux 
$m\in H^2(\T^3,\pi_1(G))$ as
well as by $c$.  As above, $m$ must be $c$-invariant.  
For all such pairs $c,m$, we expect by the same reasoning.
\eqn\pinkon{I_c(m) = (-1)^rh.}
Finally, we can include the electric flux $e$, now regarded as
a character of the $c$-invariant subgroup $\Gamma^c$ of $\Gamma$.
The  same logic leads us to predict that $I_c(e,m)=0$ unless $e$ is
a multiple of the spectral flow.  The spectral flow must be recomputed
for $c\not= 0$.

We have given two criteria for testing whether $C$ acts trivially
on the vacuum states of the infinite volume theory.  The two criteria
differ by whether the twist by $c$ is made in the ``time'' direction
or in a ``space'' direction.  They therefore are related to each
other by $SL(4,\Z)$ symmetry.  In a path integral evaluation,
the equivalence of the two criteria would be manifest. 
In section 4.5, we will test the above predictions in a microscopic
computation that is made in a Hamiltonian approach.
From this point of view, the equivalence of the two criteria is not
obvious so we will check both.
Agreement with the above predictions will
give support for the conjecture that $C$ is an unbroken
symmetry of the infinite volume theory.

\newsec{Microscopic Computations}

In this section, we begin microscopic computations to verify the predictions
made in sections 2 and 3.  In $2+1$ dimensions, we repeat part of the
analysis in \thrindex\ for completeness and as background to
$3+1$ dimensions.  In $3+1$ dimensions, the
strategy is the same as in \witten, but we will be more comprehensive.
In fact, many arguments below are given in embryonic form
in \witten\ and are presented here more fully.
In this section, we illustrate some of the ideas with simple
examples involving classical groups.  A systematic survey of the 
more elaborate examples is reserved for section 5.

\subsec{$2+1$ Dimensional Case}

We begin in $2+1$ dimensions.  We quantize on $\T^2\times \R$ the minimal
supersymmetric theory, whose Lagrangian was presented at the beginning
of section 2.  However, we specialize to $k=\pm h/2$, because
we want to            treat the low energy phase space as an orbifold;
as explained in \thrindex, this is only valid for $k=\pm h/2$.

We will analyze the low-lying states in a weak coupling approximation,
valid if the 
gauge coupling is very small or equivalently if the radius of the $\T^2$
is very small.  We begin with the case that the gauge group $G$ is 
simply-connected (as well as being simple, connected, and compact).  
We let $r$ denote the rank of $G$.

The first step is to find the classical states of zero energy.  To minimize
the energy, a configuration should be time-independent.  In addition,
the spatial part of the curvature should vanish on $\T^2$.  
This means that the restriction of the connection to $\T^2$ defines
a flat connection, which can be described up to holonomy by the monodromies
$U_1,U_2$ around the two directions in $\T^2$.  Since the fundamental
group of $\T^2$ is abelian, $U_1$ and $U_2$ commute.  

The moduli space of flat connections on $\T^2$ is the moduli space
of such commuting pairs, up to conjugation.  One can always conjugate
any one group element to lie in the maximal torus $\T_G$ of $G$.  In fact,
$G$ being simply-connected,
given any two commuting group elements $U_1$ and $U_2$, one can simultaneously
conjugate them to lie in $\T_G$.  After conjugating $U_1$ and $U_2$
into $\T_G$, one can still divide by the Weyl group $W$, so the moduli
space of flat connections on $\T^2$ is in fact
\eqn\killmop{{\cal M}=(\T_G\times \T_G)/W.}

This is a nontrivial statement whose obvious analog for three or more
commuting elements of $G$ is in general false.  
To prove that $(\T_G\times \T_G)/W$ is a component of ${\cal M}$, it
suffices to show that given $U_1,U_2\in \T_G$, any deformation of them
as a commuting pair is still (up to conjugacy) in $\T_G$.  This can be proved
by looking at the problem for first order deformations.  Such an argument
works for commuting $n$-tuples of any $n$, to show that $(\T_G\times \dots
\times \T_G)/W$ is always a component of the moduli space of commuting
$n$-tuples.

But why is $(\T_G\times \T_G)/W$ the {\it only} component of ${\cal M}$?
This is the assertion that fails in one dimension higher.
To prove the assertion in two dimensions, note that if we pick a complex
structure on $\T^2$, then ${\cal M}$ can be interpreted as 
the moduli space of holomorphic semistable $G_{\bf C}$ bundles on
$\T^2$.  Moduli spaces of semistable bundles on a Riemann surface
are always connected and irreducible.\foot{That is so because there
is no integrability condition for a $\bar\partial$ operator in complex
dimension one.  If $\bar\partial_A$ and $\bar\partial_{A'}$ are any two
$\bar\partial $ operators on the same bundle, one can literally interpolate
between them by setting $\bar\partial_t=t\bar\partial_A+(1-t)\bar\partial_{A'}$
with a complex parameter $t$.  Likewise, the moduli spaces are irreducible
since the obstruction theory is trivial.} 
There is no analog of this argument for commuting triples.

\def\M{{\cal M}}
If $G$ is not simply-connected, we should refine this description of
the moduli space slightly.
Let again $\hat G$ be the universal cover of $G$.  
 $F=\hat G/G$ is a finite abelian group that can be regarded as a quotient
 of the center of $\hat G$.   
The most important example is really that $G$ is the adjoint group (whose
center is trivial) and 
$F$ is the center of $\hat G$; considering this example gives the most
complete information.  In quantizing the
theory, according to the recipe in section 2, we impose invariance
not under the group ${\cal W}$ of all gauge transformations, but
under the restricted group ${\cal W}_0$ of gauge transformations that
are single-valued when lifted to $\hat G$.  This means that we should
consider the $U_i$ to take values in $\T_{\hat G}$, the maximal
torus of $\hat G$ (which of course is a finite cover of $\T_G$, the maximal
torus of $G$).  The phase space to be quantized is thus really
\eqn\olpo{\hat \M=(\T_{\hat G}\times \T_{\hat G})/W.}
As described in section 2, a residual group $\Gamma={\cal W}/{\cal W}_0$
acts in the problem.  An element of $\Gamma$ is a pair $\gamma=(a_1,a_2)$,
where the $a_i$ are elements of $F$ (regarded as elements of the center
of $\hat G$).  $\gamma$ acts on $\hat \M$ by
\eqn\tumilo{U_i\to a_iU_i,~i=1,2,}
and $\M=\hat\M/\Gamma$.  As we have recalled in section 2, states
of given ``electric flux'' are states in which $\Gamma$ acts with a given
character.  Until we incorporate the electric flux,
it will not matter whether the phase space is $\M$ or $\hat \M$.
The explicit
description of the moduli space in \olpo\
assumes that the magnetic flux $m$ is zero; we discuss the generalization
later.

\def\t{{\bf t}}
Now let us find the zero
modes of the fermions. A fermion zero mode must be annihilated by
the Dirac operator on $\T^2$: $\sum_{i=1}^2\Gamma^iD_i\psi=0$.
By squaring the Dirac operator and using the fact that the curvature
vanishes, we get $\sum_{i=1}^2D_i^2\psi=0$.  Upon multiplying by the complex
conjugate of $\psi$ 
and integrating by parts, we learn that $\int_{\T^2}|D\psi|^2=0$, so finally
a fermion zero mode obeys
\eqn\jcx{0=D_i\psi.}
Being covariantly constant, $\psi$ must be invariant under the monodromies
$U_1$ and $U_2$.
If we are at a generic point in ${\cal M}$, this means that $\psi$ is
a constant with values in the Lie algebra $\t$ of the maximal torus.

Now let us construct the quantum Hilbert space, in the approximation of
considering only the classical states of zero energy.  The first step
is to quantize the fermion zero modes, setting the bosons to a specific
point in ${\cal M}$.
Let $\psi_+$ and $\psi_-$ be the components of $\psi$ of positive and
negative chirality on $\T^2$.  (They are of course complex conjugates of one
another.)  The canonical anticommutation relations are 
\eqn\reggu{\{\psi_+^a,\psi_-^b\}=\delta^{ab},\,\,\,a,b=1,\dots , r.}
If we let $|\Omega\rangle$ be a state annihilated by the $\psi_-^a$, then
the fermion Hilbert space is spanned by the states 
$|\Omega\rangle,|\Omega^a\rangle=\psi_+^a|\Omega\rangle,
|\Omega^{ab}\rangle=\psi_+^a\psi_+^b|\Omega\rangle$, and so on.
We let $|\tilde\Omega\rangle$ denote the state $\psi_+^{1}\psi_+^{2}
\dots \psi_+^{r}|\Omega\rangle$ with all fermion states filled.

If we can treat $\M$ as an orbifold, then a zero energy state on 
$\M$ is the same as a zero energy state on $\T_G\times \T_G$ that
is $W$-invariant.  Taking the flat metric on $\T_G\times \T_G$, 
to get a state of zero energy, we must take the wave function to be
invariant under translations on $\T_G\times \T_G$.  The bosonic part
of the wave-function is thus a constant, which is automatically $W$-invariant.
Any state in the fermion Fock space, obtained by acting on $|\Omega\rangle$
with $s$ of the $\psi_+^a$ for any $s\leq r$, has zero energy.
The space of zero energy states is simply the $W$-invariant subspace of
the fermion Fock space (and therefore, in particular, is the same
whether we have in the quantization imposed invariance under
the restricted group
${\cal W}_0$ of gauge transformations or the full group ${\cal W}$ of
all gauge transformations).

The fermion creation operators $\psi_+^a$ transform in the representation
$\t$ of $W$.  The product of all $r$ fermion creation operators
$\psi_+^1\psi_+^2\dots\psi_+^r$ transforms in a one-dimensional real
representation $\Theta$ of $W$, in which each elementary reflection
is represented by $-1$.

It is argued in \thrindex\ that this system can be treated as an orbifold
precisely if $k=\pm h/2$, and that for those values of $k$, of
the two states $|\Omega\rangle$ and $|\tilde\Omega\rangle$, one
is $W$-invariant and the other transforms in the representation $\Theta$.
Moreover, under $k\leftrightarrow -k$, the role of the two states is
exchanged.  We will suppose that $k=\pm h/2$ is such that $|\Omega\rangle$
is $W$-invariant.

Given a complex structure on $\T^2$, the  torus $\T_G\times \T_G$
acquires a complex structure, with complex coordinates $z^a$, $a=1,\dots,r$.
The $\bar\partial $ cohomology group $H^{0,q}(\T_G\times \T_G)$
has a basis of
$(0,q$)-forms $d\bar z^{a_1}d\bar z^{a_2}\dots d\bar z^{a_q}$.
Such a form corresponds in a natural way to
$\psi_+^{a_1}\psi_+^{a_2}\dots\psi_+^{a_q}|\Omega\rangle$, which is one
of the basis elements
of the fermion Fock space, so we can think of the fermion Fock space
as
\eqn\pkoj{{\cal F}=\oplus_{q=0}^r H^{0,q}(\T_G\times \T_G).}

The space of zero energy states in the quantum theory is not ${\cal F}$
but ${\cal F}^W$, the $W$-invariant subspace of ${\cal F}$.  Imposing
$W$-invariance projects the cohomology of $\T_G\times \T_G$ to that
of $\M=(\T_G\times \T_G)/W$, so
\eqn\nkoj{{\cal F}^W =\oplus_{q=0}^r H^{0,q}(\M).}
This is the space of zero energy states.  

On the other hand, by a theorem of Looijenga
\ref\looij{E. Looijenga, ``Root Systems And Elliptic Curves,''
Invent. Math.{\bf 38} (1976) 17.} and
Bernshtein and Shvartsman \ref\bs{I. N. Bernshtein and O. V. Shvartsman,
``Chevalley's Theorem For Complex Crystallographic Coxeter Groups,''
Funct. Anal. Appl. {\bf 12} (1978) 308.} (another proof was given in
\ref\fmw{R. Friedman, J. W. Morgan, and E. Witten, ``Vector Bundles
And $F$-Theory,'' Commun. Math. Phys. {\bf 187} (1997) 679.}),
 $\M$ is a weighted projective space,
$\M={\bf WCP}^r_{s_0,s_1,\dots,s_r}$, where the weights $s_i$ are 1 and
the coefficients of the highest coroot of $G$.  (For example, for $G=SU(N)$
the weights are all 1, and for $G=E_8$ the weights are 1,2,2,3,3,4,4,5, 
and 6.)  
The $\bar\partial$ cohomology of a weighted projective space is
easily described.
$H^{0,0}(\M)$ is one-dimensional, being generated by the constant
function 1, and $H^{0,q}(\M)=0$ for $q>0$.

So  there is precisely one zero energy state,
and the index (for $m=0$) 
is $I(0)=1$, as predicted on macroscopic grounds in section 2.
Now let us include the electric flux.  Here it is important
to specify that the bosonic phase space is $(\T_{\hat G}\times
\T_{\hat G})/W$, acted on by the finite group $\Gamma$.  $\Gamma$
acts on the bosonic coordinates of the phase space according to \tumilo,
and acts trivially on the fermion zero modes.  If we fix a nontrivial
character $e$ of $\Gamma$, then the bosonic wave-function cannot
be constant and necessarily carries a strictly
positive energy.  So the one zero energy state found above
has $e=0$.  Hence, the index is as expected $I(e,0)=0 $ for $e\not=0$,
and $I(0,0)=1$.

\subsec{$3+1$ Dimensions}

Now we consider the problem of computing
the index for the $N=1$ super Yang-Mills theory in $3+1$
dimensions, formulated on a torus $\T^3$.  
In this section, we take the magnetic flux to vanish,
so the gauge field is a connection on a trivial $G$-bundle $X$.

Just as in $2+1$ dimensions, the first step is to take $g$ very small
and reduce to the moduli space of flat connections on $X$.  Such
a connection has commuting holonomies $U_1,U_2,U_3$.
It turns out, for general $G$, that the moduli space $\hat {\cal N}$ of commuting
triples has several components $\hat {\cal N}_i$.  The most obvious component,
which we will call $\hat\N_0$, 
is the one that contains the point $U_1=U_2=U_3=1$.  We will call
$\hat \N_0$ the identity component of $\hat\N$.
Any $U_i$ in $\hat\N_0$ can be simultaneously conjugated to the maximal
torus -- to be precise, the
 maximal torus of $\hat G$, since in quantization we divide only by the
restricted gauge group ${\cal W}_0$.  This component of the bosonic phase
space is then 
\eqn\ucuc{\hat \N_0=(\T_{\hat G}\times \T_{\hat G}\times \T_{\hat G})/W.}
As in $2+1$ dimensions, a finite group $\Gamma={\cal W}/{\cal W}_0$
acts on $\hat N$ by
\eqn\nutilo{U_i\to a_iU_i,~~a_i\in \hat G/G.}
The quotient $\N=\hat \N/\Gamma$ classifies flat connections
modulo the action of the full group  ${\cal W}$.
Likewise, $\N_0=\hat\N_0/\Gamma$ is the identity component of $\N$.

We will review in section 5 a systematic construction of the components
of  $\hat \N$ for any $G$,
but to orient the reader we will briefly recall a direct construction
(given in the appendix to \vectorstructure) for $G=Spin(n)$ with $n\geq 7$.
We first construct a flat $Spin(7)$ bundle $E$ on $\T^3$ that has no
moduli.  The holonomies are diagonal matrices $U_i$.  
We assume that, for $k=1,\dots,7$, the $k^{th}$ diagonal
matrix elements of $(U_1,U_2,U_3)$ are of the form
$(\pm 1,\pm 1,\pm 1)$ with each sequence of three numbers $\pm 1$
appearing precisely once except $(1,1,1)$.
(The flat bundle with these holonomies is a sum of real line bundles;
by computing Stieffel-Whitney classes, one can show that it is topologically
trivial.)
To embed this
in $Spin(n)$, we simply take the $U_i$ to be diagonal matrices
whose first seven eigenvalues are as described and whose other
$n-7$ diagonal elements are all $1$.  The subgroup of $Spin(n)$
commuting with the $U_i$ has for its connected component $Spin(n-7)$.
Let $\hat\N_1$ be the component of $\hat\N$ that contains the commuting
triple just described.  

To show that $\hat\N_1$ is distinct from $\hat\N_0$,
it suffices to observe that the unbroken groups $Spin(n)$ and $Spin(n-7)$
have different ranks.  Indeed, the rank of the unbroken subgroup
is always conserved under continuous deformation of a commuting triple.
For      if $\vec U=(U_1,U_2,U_3)$ leaves fixed a group $H_{\vec U}$,
then any small deformation of $\vec U$ as a commuting triple commutes
with a group that contains a maximal torus of $H_{\vec U}$; the small
deformation can in fact be accomplished by $U_i\to U_iV_i$ with $V_i$ in
a maximal torus of $H_{\vec U}$.
It can be shown for $Spin(n)$ that $\hat\N$ has precisely the two
components $\hat\N_0$ and $\hat\N_1$.   

For any component $\hat \N_i$,
we let $r_i$ be the rank of the unbroken group; it is one of the most
important invariants of $\hat\N_i$.  The dimension of $\hat\N_i$ is
$3r_i$, since locally  $U_1,U_2, $ and $U_3$
 vary in the maximal torus of the unbroken
group.  In particular, locally each of $U_1,$ $U_2,$ and $U_3$ is specified
by $r_i$ parameters.

\bigskip\noindent{\it Computation}

We will first describe the contribution to the index of $\hat\N_0$,
so we assume that the $U_i$ can be simultaneously conjugated to a maximal
torus.
A fermion zero mode is, as in $2+1$ dimensions, a constant
(invariant under translations on $\T^3$)
with values in the Lie algebra $\t$ of the maximal torus.
However, we get two copies of the space of fermion zero modes that we had
in $2+1$ dimensions, because the four-dimensional gluino has two
chiral components $\lambda_\alpha$, $\alpha=1,2$ and two antichiral
components $\bar\lambda_{\dot\alpha}$, $\dot\alpha=1,2$ (while the
$(2+1)$-dimensional gluino has only two real components). The zero
modes of $\bar\lambda_{\dot\alpha}$ are fermion ``creation operators''
$\bar\psi^a_{\dot\alpha}$, with $a=1\dots r$ running over a basis of
$\t$; the zero modes of $\lambda_\alpha$ are ``annihilation operators''
$\psi^a_\alpha$.  If $|\Omega\rangle$ is a ``Fock vacuum,'' 
annihilated by the $\lambda_\alpha$ modes, then
a basis
of the fermion Fock space is given by the states
\eqn\huffalo{
 \bar\psi_1^{a_1}\dots \bar\psi_1^{a_s}\bar\psi_2^{b_1}\dots\bar\psi_2^{b_t}
|\Omega\rangle,~~0\leq s,t\leq r.}

Just as in $2+1$ dimensions, the bosonic wave function of a zero energy
state must be constant,  and the space of zero energy states is simply
the Weyl-invariant part of the fermion Fock space.  The fermion
Fock space is now not simply the space ${\cal F}$ constructed
in \pkoj\ for the $(2+1)$-dimensional problem, but rather, because
we have two identical sets of fermion creation operators, it is
\eqn\unip{
 \tilde{\cal F}={\cal F}\otimes {\cal F}.}
The space of zero energy states of the $(3+1)$-dimensional gauge theory
is just  the $W$-invariant part of $\tilde {\cal F}$.

It is easy to identify $r+1$ Weyl-invariant states in $\tilde {\cal F}$.
Indeed, if 
\eqn\defrr{Z=\sum_{a,b}\delta_{ab}
\bar\psi^a_1\bar\psi^b_2,} where $\delta_{ab}$ is the
$W$-invariant metric on $\t$, then $Z$ is $W$-invariant and hence
so are the states $Z^p|\Omega\rangle$, for $p=0,1,\dots,r$.
  We want to show that there are no other $W$-invariant
states.

For this we use the following interpretation of $\tilde{\cal F}$.
As in the $(2+1)$-dimensional discussion, we consider the complex
torus $\T_{\hat G}\times \T_{\hat 
G}$ with complex coordinates $z^a$.  
\foot{
 One significant difference from the $(2+1)$-dimensional analysis
should be pointed out.  The space $\T_{\hat G}\times \T_{\hat G}$
(with precisely two factors of $\T_{\hat G}$) was a natural part of
that problem, but here has been introduced as an auxiliary tool
to get a simple description of the fermion Fock space.}
Identifying
$\bar
\psi_1^a$ as $dz^a$ and $\bar\psi_2^a$ as $d\bar z^a$, we see that
the basis of $\tilde {\cal F}$ given in \huffalo\ gives a basis
for the translationally invariant differential forms on $\T_{\hat 
G}\times \T_{\hat G}$, or in other words a basis for the de Rham cohomology
of $\T_{\hat G}\times \T_{\hat G}$.  We thus can identify 
\eqn\humbly{\tilde {\cal F}=\oplus_{p,q=0}^r H^{p,q}(\T_{\hat G}
\times \T_{\hat G}).}
In this interpretation,
 $Z^p|\Omega\rangle$ is obtained by acting on $|\Omega\rangle$
with $p$ creation operators $\bar\psi_1^a$ and $p$ creation operators
 $\bar\psi_2^a$
and hence is a form of type $(p,p)$.

The $W$-invariant part of $\tilde{\cal F}$ is hence the $W$-invariant
part of the de Rham cohomology of $\T_{\hat G}\times \T_{\hat G}$,
or in other words it is the de Rham cohomology of $\hat \N=(\T_{\hat G}
\times \T_{\hat G})/W$.  But $\hat \N$ is a weighted projective
space ${\bf WCP}^r_{s_0,s_1,\dots,s_r}$. The de Rham cohomology
of a weighted projective space of complex dimension $r$ is $(r+1)$-dimensional.
It is generated in fact by the $(p,p)$ forms $\omega^p$,
$p=0,\dots,r$, where $\omega$ is the Kahler class.  These correspond to $Z^p|\Omega\rangle$.  So
we have shown  that the space of zero energy states that come
by quantizing this component of the moduli space is $(r+1)$-dimensional.
Moreover, these states all have the same eigenvalue of $(-1)^F$, since
$(-1)^F$ commutes with $Z$.  We will analyze later the overall sign.

We should not in general conclude that (for zero magnetic flux $m$)
$\Tr(-1)^F$ 
is equal to $\pm(r+1)$.  What we have worked out is just the contribution
of the identity component $\hat\N_0$ of the moduli space $\hat\N$ of
commuting triples.
The index must
be computed as a sum of the contributions of the $\hat {\cal N}_i$.

The contributions of the other components can be worked out without
doing any essentially new computation.   
Let $\vec U=(U_1,U_2,U_3),$ be any triple of commuting elements in a component
$\hat \N_i$,
and let  $H_{\vec U}$ 
be the identity component of the 
subgroup of $G$ that commutes with the $U_i$.  The rank $r_i$ of $H_{\vec U}$
is an invariant of $\hat\N_i$, as we explained earlier.

Now, it is always possible to find in each component $\hat\N_i$
a $\vec U$  such that $H_{\vec U}$ is a simple Lie group.
(For example, in the $Spin(n)$ example considered above, we described
a point in $\N_1$ with $Spin(n-7)$ as the identity component of the
unbroken symmetry group.)
Let us call such a group $H_i$ and call its Weyl group $W_i$.
(For given $i$, there may be more than one possible $H_i$.)
If $\N_i$ were the same as $(\T_{H_i}\times \T_{H_i}\times \T_{H_i})/W_i$,
we would simply repeat the above computation and conclude that the contribution
of $\N_i$ to the index is $\pm(r_i+1)$.  This description of $\N_i$ is
not quite correct. A precise version is given in part (4) of
Theorem 1.4.1 in \bfm, and shows that $\T_{H_i}$ must be replaced
by a slightly different torus (which does not matter since the zero energy
wave functions are invariant under translation on the torus anyway)
and $W_i$ by a slightly larger group (which does not matter since the
operator $R$ introduced in \defrr\ is actually $SO(r)$-invariant,
not just $W$-invariant).  So effectively  the evaluation
of the contribution of $\hat\N_i$ to the index is equivalent to the
evaluation of the contribution of the identity component for the group
$H_i$.  
So indeed each contribution is $\pm(r_i+1)$. We will explain in section 4.3
that the signs are all equal (and we will give a precise framework
in which it is natural to set the sign to $(-1)^r$).
Hence, the index for zero magnetic flux $m$ is
\eqn\ikolo{I(0)=(-1)^r\sum_i\left(r_i+1\right)}
where the sum runs over all components of $\hat\N_i$. 
According to \refs{\newkeur,\kac,\bfm},
 this sum always equals $h$, the dual Coxeter
number of $G$, in agreement with the prediction described in section 3.

Now we want to include the electric flux $e$.  We want to show that
(still for $m=0$), the index $I(e,0)$ vanishes for $e\not= 0$.
We recall that $e$ is a character of a finite group $\Gamma$ that
acts as in \nutilo.
If we can show that the action of $\Gamma$ maps each $\hat \N_i$ to itself
(rather than permuting the $\hat \N_i$) then the argument will
go through just as in $2+1$ dimensions: a state that transforms
nontrivially under $\Gamma$ must have a nonconstant wavefunction on
$\hat\N_i$, and hence a positive energy.  If $\Gamma$ acts by nontrivial
permutations of the $\hat\N_i$, this conclusion would not follow:
in this case, suitable linear combinations of states supported on different
components would give zero energy states transforming nontrivially under
$\Gamma$, implying that $I(e,0)$ would not always vanish for nonzero $e$.

That $\Gamma$ maps each $\hat\N_i$ to itself follows from the following
facts:

(1) The commuting triple $\vec U$ defines a flat connection $A$;
the Chern-Simons invariant $CS(A)$ of this connection is $\Gamma$-invariant
mod 1.

(2) Each component of $\hat\N_i$ has a different value of $CS(A)$ mod 1.

The first assertion is essentially a consequence of the computations
in section 3.4.  Given a flat bundle on $\T^3$ defined by $\vec U$
and a gauge transformation defined by $\gamma\in \Gamma$, we can,
by using $\gamma$ as a gluing function in the ``time'' direction,
build up a flat bundle $Y$ over $\T^3\times\S^1$.  The change in $CS(A)$
under $\gamma$ is the same as the instanton number of $Y$ mod 1.
The characteristic class $\hat m\in H^2(\T^3\times\S^1,\pi_1(G))$ of $Y$
is purely electric, since we started with a bundle on $\T^3$ of zero
magnetic flux.  The computations in section 3.4 show that the instanton
number is an integer for any bundle whose characteristic class is
purely electric.  So $CS(A)$ is invariant under $\gamma$ when $m=0$.

The statement that the different $\hat\N_i$ have different values
of $CS(A)$ mod 1 is a theorem in \bfm\ (part (3) of Theorem 1.8.1).
As we will presently explain, a sharper version of this statement 
(also proved in \bfm) is actually needed
to reconcile the microscopic computation that we are presenting
here with the macroscopic arguments of section 3 concerning chiral symmetry.

Though this involves jumping ahead of our story a bit, it is convenient
to here explain the analog of the above statements
with nonzero magnetic flux $m\in H^2(\T^3,\pi_1(G))$.
Statement (2) holds for arbitrary $m$ \bfm\ (indeed,
as will be apparent, this is needed to recover from the microscopic computation
our expectations concerning chiral symmetry).  
But the computations in section 3.4 make
clear that statement (1) does not hold -- in general, $CS(A)$ is not 
$\gamma$-invariant, since the instanton number on a bundle over $\T^4$
is not always an integer.  Clearly, in order to map the individual
$\hat \N_i(m)$ to themselves (as opposed to permuting them non-trivially),
an element $\gamma\in \Gamma$ must leave $CS(A)$ invariant for bundles
of given $m$.  Statement (2) implies that the converse is also true:
$\gamma$ maps the $\hat\N_i(m)$ to themselves if and only if it leaves
$CS(A)$ fixed.   So the zero energy states are invariant not in general
under $\Gamma$ but under its subgroup $\Gamma'$ that leaves fixed $CS(A)$.
Any character of $\Gamma$ that is trivial on $\Gamma'$ is a multiple of
the spectral flow character, so we conclude that $I(e,m)=0$ unless $e$ is
a multiple of the spectral flow.

\bigskip\noindent{\it  Singularities Of The Moduli Space}

In our computations, 
we have treated the moduli spaces $\hat \N$ as orbifolds.
Actually, the description that we have        given of the low
energy effective action of the theory breaks down near singularities
of $\hat \N$.  How do we know that there are not additional
zero energy states supported near the singularities and hence invisible in our
analysis?   In what follows,  we consider $3+1$ dimensions.
(The $(2+1)$-dimensional case has been discussed in \thrindex,
and involves a few further details, revolving around
the Chern-Simons coupling.)

Suppose first  that $G=SU(2)$.
In this case, the maximal torus $\T_{\hat G}$ is just a circle
$\S$, and the Weyl group is $\Z_2$, acting on a commuting
triple $(U_1,U_2,U_3)$ (with all $U_i\in\S$) by $U_i\to U_i^{-1}$.
This transformation has eight fixed points where all $U_i$ are $\pm 1$.
At these points in the moduli space, additional bose and fermi modes,
which generically carry nonzero energy, 
move down to zero energy, and hence our description of the low energy
effective theory breaks down.  The fixed points are permuted by the
action of the finite group $\Gamma$, so they behave in the same way,
and it suffices to consider the fixed point at $U_i=1$.

It is not difficult to give a description of the low energy effective
theory that {\it is} valid near $U_i=1$.  $U_i=1$  corresponds to
 the trivial flat
connection,  $A=0$.  Expanding around $A=0$, the low energy modes are
the modes that are, in a suitable gauge, constant on $\T^3$ (but not
necessarily aligned in a Cartan subalgebra).  The effective theory
of these modes is the theory obtained by dimensional reduction
(as opposed to compactification) of the $SU(2)$ super Yang-Mills theory from
$3+1$ to $0+1$ dimensions.  In other words, the effective theory is the
$SU(2)$ ``matrix quantum mechanics'' in $0+1$ dimensions, 
with three matrices (coming from dimensional reduction of the three 
components of $A$) plus their superpartners, and with four supercharges.

\nref\fh{S.-T. Yau and J. Hoppe, ``Absence Of Zero Energy States
In Reduced $SU(N)$ 3d Supersymmetric Yang-Mills Theory,''
 hep-th/9711169.}%
\nref\setst{S. Sethi and M. Stern,  ``$D$-Brane Bound States Redux,''
hep-th/9705046.}%
\nref\pyi{P. Yi, ``Witten Index And Threshold Bound States Of
$D$-Branes,'' hep-th/9704098.}%
\nref\ssetst{S. Sethi and M. Stern, ``Invariance Theorems For
Supersymmetric Yang-Mills Theories,'' hep-th/0001189.}%
Part of the string duality picture is that matrix quantum mechanics
with 16 supercharges has zero-energy bound states, but matrix
quantum mechanics with fewer than 16 supercharges lacks them.
Absence of zero energy normalizable states with fewer than 16
supercharges has not been fully proved, but there are many partial
results \refs{\fh - \ssetst}.
So in our case, for $G=SU(2)$, we expect no zero-energy states
localized near the singularities.

If $G$ has rank higher than one, we must argue inductively.
Singularities of $\hat\N$ are loci $\hat \N_H$
on which various subgroups $H$ of $G$
are restored.  
If $G$ has rank $r$ and $H$ has rank $s$, then the relevant part of
the effective theory
near $\hat \N_H$ (but away from singularities involving groups of rank
higher than $s$) is the matrix quantum  mechanics with gauge group $H$
and four supercharges.
Given the presumed absence of zero-energy bound states in this matrix
quantum mechanics for all $H$,                                      
one would argue inductively in $s$ that none of the singular loci
support bound states.

One would also like to show that the zero-energy states that we have
seen in the orbifold quantization of $\hat \N$ do not fail to be normalizable
because of their behavior near the singularities.  For this, one must
show that in the matrix quantum mechanics, there are
supersymmetric zero-energy states that fail to be normalizable
because of their behavior at infinity on the Coulomb branch, and are in
 correspondence with the zero-energy states that we have seen on $\hat \N$.
(Infinity on the Coulomb branch in the matrix quantum mechanics
matches on to the smooth part of $\hat \N$ in the quantization of the orbifold.)
This is really a consequence of absence of normalizable zero energy
states.  Let $\psi_0$ be a smooth wavefunction that does coincide
near infinity with a supersymmetric
zero-energy state constructed on $\hat \N$ but
is not necessarily a solution of the Schrodinger equation everywhere.
Then, if $\Delta$ is the Hamiltonian of the matrix quantum mechanics,
we do not necessarily have $\Delta\psi_0=0$, but at least $\Delta\psi_0$
is an ${\bf L}^2$ state (since $\psi_0$ is a zero-energy state near infinity).
  Then we try to obey $\Delta\psi=0$ with
$\psi=\psi_0+\psi_1$ and $\psi_1$ a normalizable state (whose addition will
not modify the behavior at infinity).  The equation we have to obey
is $\Delta\psi_1=-\Delta\psi_0$, and as $-\Delta\psi_0$ is a vector
in the Hilbert space $Y$ of ${\bf L}^2$ wavefunctions, and $\Delta$ is
invertible in this space, a unique $\psi_1\in Y
$ obeying the equation
does exist.

\subsec{Chiral Symmetry Breaking}

Now we want to extend this analysis to compute the dimensions
$h^k$ of the cohomology groups $H^k(Q)$, as defined in section 3.

The key point is to compute the $R$-charge of the zero-energy vacuum
states obtained by quantizing $\hat{\cal N}_i$.  Each mode of the $\bar\lambda$
operator has charge 1.  We have to include the charge of the states
in the filled Dirac sea.  Let $N_-$, $N_0$, and $N_+$ be, formally,
the number of negative, zero, and positive eigenvalues of the 
three-dimensional Dirac operator acting on $\bar\lambda$.  Of course,
$N_-$ and $N_+$ are infinite, while, in expanding around a flat
connection $A$ in the component $\hat\N_i$, one has $N_0=2r_i$.

The charge of the Dirac sea is formally $N_-$.  To regularize
this expression, we subtract a multiple of $N=N_-+N_0+N_+$, which formally
is a constant (the total number of modes of the $\bar\lambda$ field)
and which is actually zero with zeta function regularization.
We partly regularize the charge of the Dirac sea by subtracting
$\half N$ from $N_-$, to give $\half(N_--N_+)-\half N_0
=\half(N_--N_+)-r_i$.  A regularized version of $\half(N_+-N_-)$
is given by the Atiyah-Patodi-Singer $\eta$ invariant of the flat
connection $A$, and according to the APS theorem, this $\eta$ invariant
is equal to $2h\,CS(A)$.
   The dual Coxeter number $h$ appears because
the gluino fields take values in the adjoint representation.
We therefore identify the $R$-charge of the state $|\Omega\rangle$
as $-2h\,CS(A) -r_i$.  Since $Z$ has $R$-charge 2,
the zero-energy states  $Z^s|\Omega\rangle$
have $R$-charges
\eqn\nopm{-2h\,CS(A)-r_i,-2h\,CS(A)-r_i+2,-2h\,CS(A)-r_i+4,\dots,
-2h\,CS(A)+r_i.}

In particular, for the identity component $\hat \N_0$, we can take 
$A=0$ and hence $CS(A)=0$; also the rank of $\hat\N_0$ is
 $r_0=r$.  So the $R$-charges of
these states are congruent to $r$ mod 2, and as the $R$-charge is the
fermion number mod 2, these states have  $(-1)^F=
(-1)^r$.  This is the reason for the sign choice that was claimed in
section 3.  (The values of the $CS(A)$ are such that the sign is the
same for all components \bfm; this is a special case of what we say
below in comparing to the predictions of section 3 concerning chiral
symmetry breaking.)  

The use of the APS theorem to evaluate $\eta$ assumes that there exists
a four-manifold $B$ of boundary $\T^3$ over which the connection $A$ extends.
$B$ certainly exists if the $G$-bundle $X$ on which $A$ is a connection
is topologically trivial, for then, writing $\T^3=\T^2\times \S^1$,
we can take $B=\T^2\times D$ with $D$ a disc of boundary $S^1$.
This covers the case that $G$ is simply-connected.  Even if $G$ is not
simply connected, as long as its fundamental group is cyclic, we can
always assume that the magnetic flux $\hat m$ is a pullback from $\T^2$
in some way of writing $\T^3=\T^2\times \S^1$, and then again we can take
$B=\T^2\times D$.  With more care, $B$ can be constructed also in the remaining
case $G=Spin(4n)/\Z_2\times\Z_2$. 

For any group, the Chern-Simons invariant vanishes on the identity component
$\hat \N_0$, because a flat connection with trivial holonomies
is invariant under a reflection of one axis in $\T^3$; such a reflection
reverses the sign of the Chern-Simons invariant.  A more general
result along these lines is as follows.  Suppose a component
$\hat\N_i$ of $\hat \N$ contains a point that labels the commuting
triple $(U_1,U_2,U_3)$, where $U_3^k=1$ for some integer $k$.
(As we will review in section 5, this condition is always obeyed by each
of the $\hat \N_i$.)  Then the Chern-Simons invariant of $\hat\N_i$
is an integer multiple of $1/k$.  To prove this, one considers
the four-manifold $B=\T^2\times W$ where $W$ is a $k$-holed sphere.
Because $U_3^k=1$, there is a flat connection on $B$ whose
holonomies on $\T^2$ are $U_1$ and $U_2$ and whose holonomy around any
of the boundary circles of $W$ is $U_3$.  The boundary of $B$ is $k$
copies of $\T^3$, with holonomies $(U_1,U_2,U_3)$ on each, so  we have
\eqn\huncy{k\,CS(A) ={1\over 8\pi^2}\int_B\tr F\wedge F = 0~{\rm mod} ~1.}

\bigskip\noindent{\it Examples}

Our prediction in section 3 was that there is for every value of the
$R$-charge $k$ such that $k+r$ is even a unique zero-energy state.
From \nopm, we  see that this
puts a severe restriction on the values of $CS(A)$ for the different
$\hat \N_i$ (and in particular, those values must be distinct or there
would be at least two zero-energy states of the same $R$-charge).
The fact that \nopm\ does lead after evaluation of the $CS(A)$ to
the expected spectrum of $R$-charges for the zero-energy states is
Theorem 1.8.2 in \bfm.
 Let us give some examples of how this works out for specific cases.
(See section 5 for a more elaborate example with gauge group $E_8$.)

For $G=SU(n)$, 
the dual Coxeter number is $h=n$, so the $R$-charge is defined mod $2n$.
The only component of $\hat \N$ is the identity component
$\hat \N_0$.  For this component, $r_0=n-1$, so according 
to \nopm, the values of the $R$-charge for the zero-energy states for
$SU(n)$ are
\eqn\jini{-n+1,-n+3,-n+5,\dots,n-1.}
Thus, $h^k(SU(n))=1$ if $k$ is congruent to $n-1$ modulo 2, and zero
otherwise. This is in agreement with expectations.

Similarly for $G=Sp(n)$, 
the dual Coxeter number is $h=n+1$, so the $R$-charge is defined mod $2n+2$.
$\hat \N_0$ is the only component, with
$r_0= n$.  So  the $R$-charges of zero-energy states are
\eqn\blini{-n,-n+2,-n+4,\dots,n.}
This agrees with expectations.  These examples
were described in \witten.

To give an example with more than one component, let us take 
$G=Spin(t)$.  The dual Coxeter number is $h=t-2$, so the $R$-charge
is defined mod $2h=2t-4$.
If, for example, $t=2n$ is even, then $\N_0$ has $r_0=n$ and contributes
zero-energy states of charges
\eqn\klimo{-n,-n+2,-n+4,\dots,n.}
There is one additional component $\N_1$.  A commuting triple
$(U_1,U_2,U_3)$ parametrized by a point in $\N_1$ was described
in section 4.2; it has $U_3^2=1$, so the Chern-Simons invariant for $\N_1$
is 0 or 1/2.  In fact, it is 1/2; this 
will be explained in section 5.   The rank $r_1$ is (for $t=2n$)
$n-4$.  So setting $-2h \,CS(A)=-\half(2t-4)=-2n+2$, the charges of
zero-energy states obtained by quantizing $\N_1$ are
\eqn\plimo{-3n+6,-3n+8,\dots,-n-2.}
If we combine \klimo\ and \plimo, we see (since $-3n+6$ is congruent
mod $2h=2t-4=4n-4$ to $n+2$) that there is one  zero-energy state
 for every even integer mod
$2h$, as expected based on chiral symmetry breaking from the analysis in
section 3.

The case that $t=2n+1$ is similar.  We still have $r_0=n$, so
the charges of zero-energy  states from quantization of $\N_0$ are still
given by \klimo.  But we now have $2h=2t-4=4n-2$ and $r_1=n-3$, so
the formula \plimo\ for charges of zero-energy states from quantization
of $\N_1$ is replaced by 
\eqn\newplimo{-3n+4,-3n+6,\dots, -n-2.}
Again, there is precisely one zero-energy state for every even charge
mod $2h$.

\subsec{Incorporation Of The Magnetic Flux}

We will now discuss the incorporation in this discussion of the
magnetic flux $ m\in H^2(\T^2,\pi_1(G))$ or $H^2(\T^3,\pi_1(G))$.

In $2+1$ dimensions, for any $ m$, the moduli space $\hat\M(m)$ of flat
bundles 
is always a weighted projective space \refs{\looij,\bs,\fmw}.  Hence, 
$H^{0,i}(\hat\M(m))$ vanishes except for $i=0$, and $H^{0,0}(\hat \M(m))$
is one-dimensional.  So $\Tr(-1)^F=1$ for all $ m$.

To approach this in a somewhat more intuitive way, observe that
for any $ m$,
it is possible to find a flat connection that breaks $G$ down to a simple
Lie group $H$
\ref\schwiegert{C. Schweigert, ``On Moduli Spaces Of Flat Connections
With Non-Simply Connected Structure Group,'' Nucl. Phys. {\bf B492} (1997)
743.}.
We give a simple example below.
Then the construction of zero-energy states
 for $G$ and $ m$ is equivalent
to the analysis we have already performed for gauge group $H$ and $ m=0$.
\foot{Actually,
the moduli space $\hat \M(m)$ for $G$  with magnetic flux $ m$ 
is $\T'/W'$ where
$\T'$ is a slightly different torus from $\T_{\hat H}\times \T_{\hat H}$ and
$W'$ is a slightly larger group than $W_H$.   See Theorem 1.3.1 of
\bfm\ for a precise
statement.
Neither of these points affects the identification of zero-energy states
since the zero-energy states are constants on $\T'$ anyway, and
 the only  $W_H$-invariant state in the quantization has full $SO(r)$ 
symmetry.}
So since $\Tr\,(-1)^F=1$ for all $H$ at $\hat m=0$, it equals 1 for
all $G$ and all $ m$.

Now, to complete the verification of the
predictions in section 2,
let us show, still in $2+1$ dimensions,
 that $I(e,m)=0$ for all nonzero $e$.
First of all, a flat connection on a $G$-bundle with $m\not = 0$ can
be explicitly described by a pair of monodromies $U_1,U_2$ that commute
in $G$            but which, if lifted to the universal cover $\hat G$ of
$G$, do not quite commute.  Instead, as in \juffy, they obey
\eqn\jujugo{U_1U_2=U_2U_1 \hat f,}
where $\hat f$ is a central element of $\hat G$ that is determined
by $m$.  Thus $U_1$ and $U_2$ are ``almost commuting.''  
The residual symmetry group $\Gamma$, of which the electric flux $e$ is
a character, acts as in \tumilo\ by
\eqn\pujugo{U_i\to a_i U_i}
with $a_i$ central elements of $\hat G$.  In particular, $\Gamma$ always
acts by translations on the appropriate torus
and hence leaves invariant the only zero
energy state.  This leads to our claimed result that $I(e,m)=0$ for
$e\not= 0$, and $I(0,m)=1$.

\bigskip\noindent{\it $3+1$ Dimensions}

In $3+1$ dimensions, things are slightly more elaborate.
First of all, for $m\not= 0$, just as for $m=0$, there are
several components $\hat \N_i(m)$ of the moduli space $\hat \N(m)$.
On each of these, the unbroken subgroup of $G$ is
of rank $r_i$ for some $r_i$.  At some
point on the moduli space, the unbroken group is a simple   group $H_i$
of rank $r_i$ (more than one $H_i$ may appear at different points on the
same $\hat\N_i(m)$).  
The contribution of $\hat\N_i(m)$ to $\Tr\,(-1)^F$ is hence
$\pm (r_i+1)$.  
To get the expected result $\Tr\,(-1)^F=(-1)^rh$, we hence need
\eqn\importo{\sum_i(r_i+1)=h,}
and moreover the signs should agree.
The assertion \importo\ has been proved in Theorem 1.5.1 of \bfm, 
and the assertion about the signs is a special case of
 Theorem 1.8.2 of that paper, which asserts that the
values of the Chern-Simons invariants of the various components of $\hat
\N_i(m)$ are such that there is precisely one zero-energy state for
every even or every odd integer mod $2h$. 

As for the further prediction that $I(e,m)=0$ unless $e$ is a multiple
of the spectral flow, the reason for this has already been explained
in section 4.2: an element of $\gamma$ leaves fixed the individual
$\hat\N_i(m)$ if and only if it leaves fixed the Chern-Simons invariant
of a bundle of magnetic flux $m$.

\bigskip\noindent{\it Specialization To $SU(n)/\Z_n$}

In the rest of this section, we examine in detail how these predictions
work out for $G=SU(N)/\Z_n$.
For details concerning the other classical groups, see  section 5.3.

Since the fundamental group of $G$ is the cyclic group $\Z_n$, we can
pick a way of writing $\T^3=\T^2\times \S^1$ such that $m$ is a pullback
from the first factor.  Let us first look at the basic example that the
magnetic flux on $\T^2$ is ``1.''  This means that the holonomies
$U_1,U_2,U_3$ of a flat connection, if lifted to $SU(n)$, obey
\eqn\kicox{U_1U_2=U_2U_1\exp(2 \pi i/n)}
and
\eqn\nicox{U_iU_3=U_3U_i~{\rm for}~i=1,2.}
$SU(n)$ elements $U_1$, $U_2$ obeying \kicox\ exist and are unique up
to conjugation.  Moreover, such a $U_1$ and $U_2$ commute only
with the center of $SU(n)$ so 
\eqn\jucent{U_3=\exp(2\pi i p/n)} for some integer $p$.
The moduli space $\hat \N(m)$, for this $m$, thus consists of
$n$  points, with $U_1$ and $U_2$  given by some fixed solution of
\kicox\ and $U_3$ as in \jucent.  

Each such point is a connected
component $\hat\N_p(m)$ of $\hat \N(m)$, for $p=1,\dots,n$.  
The expected formula
$\sum_p(r_p+1)=h$ is obeyed  with $p$ taking $n$ possible values,
$r_p=0$ for all $p$, and $h=n$. 

Now let us study the action of $\Gamma$.  An element $(a_1,a_2,a_3)$
of $\Gamma$, with each $a_i$ being an element of the center of $SU(n)$,
say \eqn\hurot{a_i=\exp(2\pi i \alpha_i/n),~~~\alpha_i\in\{0,1,\dots,n-1\},}
 acts on
the $U_i$ by 
\eqn\humbo{U_i\to a_iU_i.}
The action of $a_1$ and $a_2$ is trivial; $U_1$ and $U_2$ are mapped
to a different solution of \kicox, but any two such solutions are
equivalent up to conjugation.  The action of $a_3$ permutes the
$\hat\N_p$ by $p\to p+\alpha_3$.
The condition  for $\gamma$ to leave fixed the individual $\hat\N_p$ is
thus precisely that $\alpha_3=0$.
Next, let us find the condition for $\gamma$ to leave fixed the Chern-Simons
invariant of a connection on $\T^3$.  Given a connection $A$ on a $G$-bundle
$X$ over $\T^3$,
we use $\gamma$ as a gluing function in the time direction to build
a $G$-bundle $\hat X$ over $\T^4=\T^3\times \S^1$.  
The change under $\gamma$ in the Chern-Simons invariant 
of $A$ is the instanton number of $\hat X$ mod 1.
The bundle
$\hat X$ has a characteristic class $\hat m$ whose magnetic part
(the restriction to $\T^3$) is the original $m$, while the electric
part is determined by $\gamma$.  As we have computed in \hch, the instanton
number mod 1 of $\hat X$ is 
\eqn\inmodone{\Delta'(m,\gamma)=
{\Pf(\hat m)\over n}={\hat m_{12}\hat m_{34}+\hat m_{23}\hat m_{14}
+\hat m_{31}\hat m_{24}\over n}.}
We have $\hat m_{i4}=\alpha_i$ for $i=1,2,3$, and 
$\hat m_{ij}=m_{ij}$ for $i,j=1,2,3$.  Finally, the only
nonzero $m_{ij}$ is $m_{12}=1$.  So we get
\eqn\pinmodone{\Delta'(m,\gamma) = {\alpha_3\over n}.}
So the condition to have $\Delta'(m,\gamma)=0$ is $\alpha_3=0$,
which, as we aimed to prove, is the same as the condition for $\gamma$
to leave fixed the individual $\hat\N_p$. 

Since the Chern-Simons invariant of $\hat\N_0(m)$ is zero, 
 the Chern-Simons invariant of $\hat\N_p(m)$ is equal to the change
in Chern-Simons invariant
 under a gauge transformation with $\alpha_3=p$ and hence is
\eqn\todays{CS(A)={p\over n}.}
The state obtained by quantizing $\hat\N_p(m)$ hence has
$R$-charge $-2h\,CS(A)=-2n(p/n)=-2p$, and so, taking all values of $p$,
there is precisely one zero-energy state of every even charge mod $2n$,
as expected.

It is not difficult to extend this to an arbitrary $m$, still keeping
$G=SU(n)/\Z_n$.  There is no essential loss in assuming that the only
nonzero component of $m$ is $m_{12}$, which we take to be $q\in \Z_n$.
The conditions for a flat connection are now
\eqn\vucu{U_1U_2=U_2U_1\exp(2\pi i q/n)}
as well as
\eqn\blucu{U_iU_3=U_3U_i,~{\rm for }~i=1,2.}
Let $u$ be the greatest common divisor of $n$ and $q$, and $v=n/u$.  
Then \vucu\ has
an irreducible solution   in $v\times v$ matrices $A_1,A_2$.  
We embed $SU(u)\times SU(v)$ in $SU(n)$ such that the ${\bf n}$ of
$SU(n)$ decomposes under $SU(u)\times SU(v)$ 
as ${\bf u}\otimes {\bf v}$.  The  general
solution   of \vucu\ and \blucu\ is up to conjugacy
\eqn\umucu{\eqalign{U_1 & = M_1\otimes A_1 \cr
                    U_2 & = M_2\otimes A_2  \cr
                    U_3 & = M_3\otimes \exp(2\pi i p/n),\cr}}
where the $M_i$ are commuting elements of $SU( u)$ and $p\in \{0,
1,\dots,v-1\}$; we identify $p\cong p+v$ since a factor
$\exp(2\pi i /u)$ can be absorbed in $M_3$.
  The moduli
space $\hat \N(m)$ has $v$ components $\hat\N_p(m)$, labeled by
$p$ which appears in \umucu, and each component has unbroken
subgroup of rank $r=u-1$.  The sum $\sum_i(r_i+1)$ thus equals
$uv=n$, as expected.

By repeating the above computation, we see that
 $\gamma=(a_1,a_2,a_3)$ changes the instanton number by
\eqn\gumuc{\Delta'(m,\gamma)={q\alpha_3\over n}~ {\rm mod} ~ 1.}
The condition for this to vanish is that $\alpha_3$ should  be divisible
by $v$.  $\gamma$ acts on $p$ by $p\to p+\alpha_3$, so likewise,
$\hat\N_p(m)$ is mapped to itself if and only if $\alpha_3$ is divisible
by $v$.   \gumuc\ also means, since $\hat\N_0(m)$ has Chern-Simons
invariant zero, that $\hat\N_p(m)$ has Chern-Simons invariant
\eqn\ikni{CS(A) = {qp/n},}
which, as $p$ varies, ranges over all integer multiples of $1/v$.
From this, one can again show that there is one zero-energy state of
every even or every odd $R$-charge.

\subsec{Outer Automorphisms}

Finally, we wish to give some examples of verifying the predictions
given in section 3.5 for gauge groups $G$ with a group $C$ of
outer automorphisms.  As we recall, $C$ is realized as a global
symmetry group, and we want to check the claim that the vacuum states
of the infinite volume theory are all $C$-invariant.

We gave two criteria in section 3.5.  One is to compute
$I(c)=\Tr\,c(-1)^F$.  The prediction is that it is independent of $c$.
To verify this, it suffices to check that the vacuum states found in
the above computation are all $c$-invariant.  Then, computing the trace
as a sum over these states, we get a $c$-invariant result.

To prove that the vacua are $C$-invariant, the key point (rather as
in the analysis of electric flux) is to show that each component
of $\N$ is mapped to itself by $C$ (as opposed to $C$ permuting
the components).  Indeed, the Chern-Simons three-form of a 
connection is defined using a quadratic form on the Lie algebra
that is both $G$-invariant and $C$-invariant, so in particular
the Chern-Simons invariant of a flat connection is $C$-invariant.
Since the different components  $\N_i$ of $\N$ have different Chern-Simons
invariants, each is mapped to itself by $C$.

$C$ may act nontrivially on both the boson and fermion zero modes
that are encountered in quantizing $\N_i$. The action on the boson
zero modes is irrelevant, since the zero energy states are independent
of the bosons.  The action on the fermions comes from an orthogonal
transformation on the Lie algebra index of the fermion zero
modes $\lambda^a_\alpha$, $\bar\lambda^a_{\dot\alpha}$.
The fermion bilinear $Z$ introduced in \defrr\ is not just Weyl-invariant
but invariant under all orthogonal transformations of the Lie algebra
of the maximal torus, and hence in particular is $C$-invariant. 
So the zero energy states are all $C$-invariant, as expected.

\bigskip\noindent{\it The Second Criterion}

The second criterion for $C$ to be unbroken involved introducing
 a bundle $X_c$ twisted by an element of $c$ and computing
the index $I_c$ for the supersymmetric theory quantized on such a bundle.
We will verify the
predictions in several representative cases.

First we consider the case that $G=SU(n)$ and $G'$ is an extension
of $G$ by the complex conjugation automorphism $c$.  Thus, there is
an exact sequence
\eqn\uncu{1\to G\to G'\to C\to 1,}
where $C=\Z_2$ is generated by $c$.

We work on the $G'$ bundle $X_c$ over  $\T^3$ described in section 3.5.
This means that, writing $\T^3=\S^1\times \T^2$, the fields are
conjugated by $c$ in going around the $\S^1$.  
A flat connection on $X_c$ has commuting holonomies $(U_1,U_2,U_3)\in G'$
whose images in $C$ are $(c,1,1)$.  There are two components of the moduli
space $\hat \N$
of commuting triples.  One component $\hat\N_0$ has a representative
with $(U_1,U_2,U_3)=(c,1,1)$; the unbroken group $H_0$ is $SO(n)$.
The second component $\hat\N_1$ has a representative with
$U_1=c$, $U_2={\rm diag}(-1,-1,1,1,\dots,1)$, $U_3={\rm diag}(1,-1,-1,1,
1,\dots,1)$.  The unbroken subgroup is $H_1=SO(n-3)$.  If we let
$r_i$ denote the rank of $H_i$ for $i=0,1$, then the expected formula
$\sum_i(r_i+1)=h$ becomes $(r_0+1)+(r_1+1)=n$.  This holds for all $n$,
even or odd.

Now let us consider $G=Spin(n)$, and let $G'$ be the extension
of $G$ by a reflection $c$ of one of the coordinates in the $n$-dimensional
vector representation of $Spin(n)$.  Thus, in that representation,
we can think of $c$ as the diagonal matrix $c={\rm diag}(-1,1,1,\dots,1)$.
We again work on a bundle   twisted by $c$ in going around the first
factor in $\T^3=\S^1\times\T^2$.
The moduli
space $\hat \N$ of commuting  triples $(U_1,U_2,U_3)$ in $G'$
that map to $(c,1,1)$ in $C$
has again two components.   One component $\hat\N_0$
has a representative    with $U_1=c$, $U_2=U_3=1$,
and unbroken group $H_0=SO(n-1)$.  To describe the second component,
note that for $n=6$, one can make a commuting triple that has the desired
image in $C$ by taking the $U_i$ to equal diagonal matrices
$V_i$ whose diagonal elements are all triples of the numbers
$\pm 1$ except $(1,1,1)$ and $(-1,1,1)$.  Then if we set $U_i=V_i\oplus 1$,
with $1$ the identity element of $SO(n-6)$, we get a commuting triple
in a component $\hat\N_1$ with $H_1=SO(n-6)$.  The relation
$\sum_i(r_i+1)=h$ becomes $(r_0+1)+(r_1+1)=n-2$, and this holds
for all $n$, even or odd.

\newsec{More On Commuting Triples}

In this section, we will give more information about commuting
triples in a simple Lie group $G$.
Following some of the ideas in \refs{\newkeur,\kac,\bfm} as well
as comments by A. Borel, R. Friedman and J. Morgan,  we will explain
the formula $\sum_i(r_i+1)=h$, where the sum runs over the components
of the moduli space of commuting triples.
The goal is not to give proofs (which can be found in the references) 
but to present some facts that some readers may find helpful. 
We will also explain how to concretely construct a commuting triple
in each component, and how to compute its Chern-Simons invariant.

\subsec{Simply Connected Case}

We assume first that the compact, simple Lie group $G$ is also
connected and simply-connected.  

The moduli space ${\cal M}$ of commuting pairs $U_1,U_2$ in $G$ is
connected.  For a generic point $\vec U=(U_1,U_2)$ in ${\cal M}$,
the subgroup $H_{\vec U}$ of $G$ consisting of elements that commute
with the $U_i$ is connected.  

Consider a commuting triple $(U_1,U_2,U_3)$, and let $\vec U$ still
denote the pair $(U_1,U_2)$.  
Obviously, $U_3\in H_{\vec U}$,
so if
$H_{\vec U}$ is connected, we can continuously deform  $U_3$ to the
identity while preserving the fact that it commutes with $U_1$ and $U_2$.
Then, since ${\cal M}$ is connected, we can continuously deform $U_1$ and $U_2$
to the identity while preserving the fact that they commute.

So in short, a commuting triple $(U_1,U_2,U_3)$ can represent a point
in a non-identity component of the moduli space $\N$ of commuting
triples only if $H_{\vec U}$ is disconnected and $U_3$ is in a non-identity
component of $H_{\vec U}$.  

\def\C{{\bf C}}
It turns out that the moduli space ${\cal M}$ of commuting pairs
is an orbifold,
and has orbifold singularities precisely at those points at which
$H_{\vec U}$ is disconnected.  This is somewhat surprising, as one
might expect a singularity whenever $H_{\vec U}$ (which generically is
abelian) becomes nonabelian, whether it is connected or not.
But it turns
out that a singularity in the complex structure of ${\cal M}$
only arises if $H_{\vec U}$ is disconnected.
The group of components of $H_{\vec U}$ is always a cyclic group $\Z_k$
for some $k$, and the singularity of ${\cal M}$ is an orbifold singularity
$\C^r/\Z_k$, with $r$ the rank of $G$ and some linear action of $\Z_k$ on
$\C^r$.

In fact, ${\cal M}$ is always a weighted projective space
${\bf WCP}^r_{s_0,s_1,\dots,s_r}$ with the weights being
1 and the coefficients of the highest coroot of $G$ \refs{\looij,\bs,
\fmw}.  In particular, one always has 
\eqn\juniper{\sum_i s_i=h.}
We can describe ${\bf WCP}^r_{s_0,s_1,\dots,s_r}$ via homogeneous
coordinates $z_i$ of weight $s_i$.  The moduli space is the quotient
of $\C^{r+1}$ minus the origin by $\C^*$, where $\C^*$ acts by
\eqn\turnin{z_i\to \lambda^{s_i}z_i,~~ {\rm for} ~\lambda\in \C^*.}
An orbifold singularity arises if for some set of $z_i$, not all zero,
one has $z_i=\lambda^{s_i}z_i$.  This happens for $\lambda=\exp(2\pi i u/k)$
(with relatively prime integers $u$ and $k$) precisely
if $z_i=0$ unless $k$ is
a divisor of $s_i$.

Before explaining the general picture, let us work out what happens
for the classical groups.
For $G=SU(n)$ or $Sp(n)$, the weights are all $s_i=1$.
${\cal M}$ is an ordinary projective space, and in particular is smooth.
So $H_{\vec U}$ is always connected, and ${\cal N}$ has only the
identity
 component ${\cal N}_0$.

Now let us consider $G=SO(2n)$. ($SO(2n+1)$ behaves quite similarly.)
The weights are $1,1,1,1,2,2,\dots,2$ with four $1$'s and the rest $2$'s.
An orbifold singularity arises precisely if $z_1=\dots =z_4=0$.
It is a $\Z_2$ orbifold singularity.  On this locus, $U_1$ and $U_2$
are parametrized by $n-4$ complex parameters (the remaining $z_i$
modulo scaling by $\C^*$), so the unbroken group has rank $n-4$.
So for $SO(2n)$, $\N$ has the identity component $\N_0$, of rank $r_0=n$,
and a second component $\N_1$, of rank $r_1=n-4$, confirming that
the components considered in section 4.2 are the only ones.

Now let us work out the case of a general $G$.  We fix an integer
$k$ which is a divisor of some of the $s_i$, and we let $\mu(k)$ be the
number of $s_i$ that are divisible by $k$.  Then $\mu(k)-1$ is the 
complex dimension of the subspace $\M_k$ of $\M$ on which there is a $\Z_k$
orbifold singularity.  Hence the rank $r(k) $ of a component of $\N$
for which $U_1,U_2\in \M_k$ is $r(k)=\mu(k)-1$.  For $U_1,U_2
\in\M_k$,
the unbroken group $H_{\vec U}$ has a group of components that is
$\Z_k$.  $U_3$ could lie in any component of $H_{\vec U}$, but to avoid
multiple-counting, we want to consider only
 components of $\N$ that could not be constructed using a smaller
value of $k$.  For this,
 we assume that $U_3$ lies in a component of $H_{\vec U}$
whose image in $\Z_k$ is a generator of $\Z_k$; the number of such components
is $\phi(k)$, the number of integers
 mod $k$ that are prime
to $k$.

The contribution to $\sum_i(r_i+1)$ from a component of $\N$ associated
with a $\Z_k$ orbifold singularity in $\M$ is hence $\phi(k)(r(k)+1)
=\phi(k)\mu(k)$.  Since $h=\sum_is_i$, to prove that $\sum_i(r_i+1)=h$,
we need
\eqn\olono{\sum_k\phi(k)\mu(k) =\sum_is_i.}
In fact, for each $s_i$, one has the elementary number theory formula
\eqn\polono{\sum_{k|s_i}\phi(k)=s_i.}
Summing \polono\ over $i$ and using the definition of  $\mu(k)$ as the
number of $i$ such that $k|s_i$, we arrive at \olono.

\bigskip\noindent{\it Explicit Construction}

Now we want to give some indications of how to explicitly describe
a commuting triple in each component and compute the Chern-Simons invariants.

If one selects an integer $k$ and omits from the extended Dynkin diagram
of $G$ all nodes whose label $s_i$ is not divisible by $k$, one is left
with the Dynkin diagram of a Lie group $W$ of rank $r-\mu(k)$.
A fact that  is not obvious but is easy to check by examining all
the examples (using the classification of simple Lie groups and their
extended Dynkin diagrams) is that this group
is always a product $W=SU(w_1)\times SU(w_2)\times \dots \times SU(w_s)$
of $SU$ groups of various ranks.  (A more conceptual explanation is in \bfm.)

This does not mean that $W$ is a subgroup of $ G$.  Rather,
$  G$ contains a subgroup $W/D$, where $D$ is a cyclic subgroup
of the center $Q=\Z_{w_1}\times\Z_{w_2}\times\dots\times \Z_{w_s}$ of $W$.
$D$ always has the property that, if one omits all factors in $Q$ but one,
$D$ projects surjectively onto  that factor.

Hence if $f$ is a generator of $D$, and $U_1,U_2\in W$ are such that
\eqn\tomigo{U_1U_2=U_2U_1f,}
then the subgroup of $W$ that commutes with $U_1$ and $U_2$ is only
the center.\foot{In any given $SU(w)$ factor of $W$, the equation reads
$U_1U_2=U_2U_1\exp(2\pi i p/w)$, where $p$ is prime to $w$.  This follows
from the fact that $D$ projects surjectively onto the center of $SU(w)$.}
In $ G$, $U_1$ and $U_2$ commute.  For any $f'\in Q$,
$(U_1,U_2,f')$ is a commuting triple.  For suitable $f'$, we get
the commuting triples described above.

Since $(U_1,U_2,f')$ can be regarded as elements of $W$, which is
a product of $SU$ groups, the Chern-Simons
invariant of this commuting triple can be evaluated using the explicit
computation in section 4.4 for $SU(n)$.  This is analogous to what
was done in section 3.4, where the evaluation of the spectral flow for
any $G$ was reduced to a computation for $SU(n)$.

Let us carry out this procedure for an illustrative example.
We set $ G=E_8$ and $k=5$.  There is a single node on the $E_8$
Dynkin diagram of weight divisible by (and in fact equal to) 5; 
if we omit it, we are left with the Dynkin diagram of
$W=SU(5)\times SU(5)$, whose center is $Q=\Z_5\times \Z_5$.
The decomposition of the adjoint representation
of $E_8$ under $W$ is 
\eqn\julko{{\bf 248}=({\bf 24},{\bf 1})\oplus ({\bf 1},{\bf 24})
\oplus ({\bf 5},{\bf 10})\oplus (\bar{\bf 5},\bar{\bf 10})\oplus
({\bf 10},{\bar {\bf 5}})  \oplus(\bar{\bf 10},{\bf 5}).}
The subgroup $D$ of $Q$ that acts trivially on the ${\bf 248}$ is
generated by $f=\exp(2\pi i/5)\times \exp(4\pi i/5) \in SU(5)\times SU(5)$.
The solution $U_1,U_2$ of \tomigo\ is unique up to conjugation.
Upon setting $f'=\exp(2\pi ib/5)\times 1$ with $b\in\{1,2,3,4\}$,
we get a commuting triple $(U_1,U_2,f')$.  
This triple completely breaks $E_8$ and hence has rank zero,
as expected since only a single weight on the $E_8$ Dynkin diagram
is divisible by 5.  To verify that the triple completely breaks $E_8$,
it suffices to note that $f'$ breaks $E_8$ to $SU(5)\times SU(5)$ 
(as is clear from \julko) and that $U_1$ and $U_2$ completely break
$SU(5)\times SU(5)$.

To compute the Chern-Simons invariant, we just have to compute
the Chern-Simons invariant of the triple $(U_1,U_2,f')\in SU(5)\times SU(5)$.
Since $f'$ is in the first factor, this is the same as the Chern-Simons
invariant of $(U_1,U_2,\exp(2\pi i b/5))$ in $SU(5)$, and according to 
\todays\ is $b/5$.  While we have explained this for a particular $G$ and 
a particular $k$, in general,
the  Chern-Simons invariants of all of the components
of $\hat\N$ can be computed in this way. 

For components of $\hat\N$ built starting with a divisor $k$,
the Chern-Simons invariants turn out to be, by a computation similar
to the case 
that was just explained, $b/k$ where $b$ ranges over 
the integers prime to $k$ mod $k$.
With this information and the ranks $r(k)=\mu(k)-1$, the $R$-charges
of the zero-energy states can be computed.
As stated in Theorem 1.8.2
of \bfm, this leads to a spectrum of 
$R$-charges that agrees with the expectations 
based on chiral symmetry breaking.
The results for $E_8$ are summarized 
in the table.  For other simply-connected
groups, the results are similar though (since fewer values of $k$ enter)
perhaps less elaborate.  The results for classical groups were described
in section 4.3.

\bigskip
\bigskip
\tolerance = 10000
\hfuzz = 5pt

\input tables
\begintable
~~~$k$~~~ | ~~~Rank ~~~| ~~~CS Invariants ~~~| ~~~$R$-Charges of Vacua~~~\cr
1 | 8 | 0 | ~~~ $-8$, $-6$, $-4$, $-2$, 0, 2, 4, 6, 8~~~\cr
2 | 4 | 1/2 | 26, 28, 30, 32, 34\cr
3 | 2 | 1/3, 2/3 | 18, 20, 22, 38, 40, 42\cr
4 | 1 | 1/4, 3/4 | 14, 16, 44, 46\cr
5 | 0 | 1/5, 2/5, 3/5, 4/5 | 12, 24, 36, 48\cr
6 | 0 | 1/6, 5/6 | 10, 50\endtable

\noindent
This table shows the divisors 
$k$ for $E_8$, the ranks 
of the associated components 
of ${\cal N}$, their Chern--Simons (CS) 
invariants, and the $R$-charges 
of zero energy vacuum states obtained by quantizing them.  
Every even integer mod $2h = 60$ appears precisely once in the list of
$R$-charges.
\bigskip
\bigskip

For one more example, take $G=Spin(2n)$ and $k=2$.  Omitting all nodes
of the extended Dynkin diagram that have weight divisible by (and in fact
equal to) 2, we are left with the Dynkin diagram of $W=SU(2)^4$.  The
subgroup of $G$ is in fact $W/\Z_2$.  (The nontrivial element of $\Z_2$ is
the product of the elements $-1$ in each of the four factors of $W$.)
In the above construction, one can take $f'$ to be the element $-1$ of
the first $SU(2)$ factor in $W$.  The embedding of $W/\Z_2$ in $Spin(8)$ uses
the embeddings (here stated at the Lie algebra level)
 $SU(2)\times SU(2)=Spin(4)$
and $Spin(4)\times Spin(4)\subset Spin(8)$.
Via this chain, the elements $(U_1,U_2,U_3)$ constructed as above
can be regarded as elements of $Spin(8)$.  Indeed, following
through the definitions, one sees that the $U_i$ are diagonal
matrices whose eigenvalues are all possible triples of numbers
$\pm 1$.  When this is embedded in $Spin(2n)$,  we get the description given 
in section 4.2.  The Chern-Simons invariant is 1/2; the computation can
be done in the first $SU(2)$ factor of $W$ (since $f'$ is in this factor)
and equals $1/2$ by virtue of \todays.

\subsec{ Inclusion of Magnetic Flux}

Now we want to repeat this for $G$ that is connected but not
necessarily simply-connected.  
The novelty now is that one can include
the magnetic
flux $m\in H^2(\T^3,\pi_1(G))$ and the dual electric flux $e$.
We may as well assume that $G$ is the adjoint group (with trivial center);
picking a different $G$  amounts to restricting the choices of $m$ and $e$.

As in section 4, we let $\hat G$ be
the universal cover of $G$, and $\hat \N$ the moduli space of flat
$G$ connections modulo gauge transformations that are single-valued
if lifted to $\hat G$.  On $\hat \N$ acts a finite group
$\Gamma={\rm Hom}(\pi_1(\T^3),\pi_1(G))$; the quotient is
$\N=\hat\N/\Gamma$.
Let $\Gamma_0$ be the subgroup of $\Gamma$ that leaves fixed
the Chern-Simons invariant.

As we have discussed in section 3, when $\pi_1(G)\not= 0$, we can
formulate the idea of confinement and oblique confinement.  It is
believed that for all $G$, the $N=1$ supersymmetric Yang-Mills theory
with gauge group $G$ is confining.  As for whether the confining vacua have
``ordinary'' or ``oblique'' confinement, we have surveyed the possibilities
in section 3.4. For each $G$ and $m$, there is a ``spectral flow''
character $\Delta'(m,\gamma)$ computed in section 3.4. 
It is defined as the change in the Chern-Simons invariant of a bundle of
flux $m$ under a gauge transformation $\gamma$.
Let $w(G,m)$
be the order of $\Delta'$ (in the finite group $E$ of characters of 
$\Gamma$ or electric fluxes).
$\gamma\in\Gamma$ lies in $\Gamma_0$  if and only if $\Delta'(m,\gamma)=0$.
So $w(G,m)$ is the index of $\Gamma_0$ in $\Gamma$.

For simplicity, we assume that $\pi_1(G)$ is cyclic (this omits
only the case $G=SO(4n)/\Z_2$).  In that case, as we have discussed
in section 3, for some way of writing $\T^3=\T^2\times \S^1$,
$m$ is a pullback from the first factor.

A flat connection on $\T^3$ with flux $m$ can be described by commuting
holonomies $U_1,U_2,U_3$ that if lifted to the universal cover $\hat G$
obey
\eqn\unic{\eqalign{U_1U_2&=U_2U_1f \cr
                   U_iU_3&=U_3U_i ~~{\rm for}~i=1,2.\cr}}
Here, as explained in section 2.1,
 $f$ is an element of the center of $\hat G$ determined by  $m$.

For given $\vec U=(U_1,U_2)$, $U_3$ must lie in the subgroup $H_{\vec U}$
consisting of elements of $\hat G$ that commute with $U_1,U_2$.
The difference from the previous case is that in general, $H_{\vec U}$ is
not connected for generic $U_1,U_2$.  The number of components is
precisely $w(G,m)$.  
In fact, consider an element of $\Gamma$ of the form $\gamma=(1,1,a)$, with
$a$ an element of the center of $\hat G$.  If $a$ is in the identity
component of $H_{\vec U}$, then the  triple $(U_1,U_2,a)$
can be continuously deformed (preserving \unic) to $(U_1,U_2,1)$
and hence $\gamma$ leaves fixed the Chern-Simons invariant.  So the
subgroup of the center of $\hat G$ that is in the identity component of
$H_{\vec U}$ is of index at least $w(G,m)$, and hence $H_{\vec U}$ has at least
$w(G,m)$ components.  This is the actual number of components, for
given $U_1,U_2$.  For instance, in an example considered in section 4.4,
with $G=SU(n)/\Z_n$ and $m$ equal to ``1,''
we have $w(G,m)=n$, and $H_{\vec U}$ consists precisely of
the center of $SU(n)$, which consists of $n$ points, each of which is
a connected component.

Even when $G$ is not simply-connected,
the moduli space $\hat\M$ of flat $G$-bundles on $\T^2$ with
magnetic flux $m$ is still a weighted projective space
${\bf WCP}^t_{u_0,u_1,\dots,u_t}$ for some $t$ and some weights $u_i$.
It is no longer true that $t$ equals the rank of $G$, or that
$\sum_i u_i=h$.  Rather, 
\eqn\huffo{h=w(G,m)\sum_iu_i.}
(For instance, in the above-cited  example for $SU(n)/\Z_n$, we have
 $h=w(G,m)=n$, $t=0$, and $u_0=1$.)

Just as before, if we set to zero all homogeneous coordinates whose
weights are not divisible by some integer $k$, we get a locus of
orbifold singularities in $\hat\M(m)$.  On this locus, the number of components
of $H_{\vec U}$ is $w(G,m)k$.  Repeating the
computation that led to \olono, the contribution to $\sum_i(r_i+1)$
from orbifold singularities of order $k$ in $\hat \M(m)$ is
not $\phi(k)\mu(k)$, as we had before,
but $w(G,m)\phi(k)\mu(k)$.  Using \polono\ and
\huffo, we hence get $\sum_k w(G,m)\phi(k)\mu(k)=w(G,m)\sum_iu_i=h$,
as desired.

\subsec{Examples With Nonzero $m$}

We conclude with a survey of some examples with nonzero $m$.
The discussion is in rough parallel with the evaluation of the spectral
flow for various examples in section 3.4.  We have already analyzed
$SU(n)$ in section 4.4.  We consider here the other classical groups.
Except for $Spin(4n)$, we can assume (as shown in section 3)
that the magnetic flux has only one nonzero component $m_{12}$.

\bigskip\noindent{$Sp(n)$}

The center of $Sp(n)$ is $\Z_2$, so there is only
one possible nonzero value of $m_{12}$.  A flat $Sp(n)/\Z_2$ bundle
with $m_{12}\not= 0$ has holonomies that if lifted to $Sp(n)$ obey
\eqn\icono{\eqalign{U_1U_2 & = -U_2U_1  \cr
                      U_i U_3 & = -U_3 U_i,~{\rm for}~i=1,2.\cr}}

We use the embedding $Sp(1)\times O(n)\subset Sp(n)$, under which
the fundamental representation of $Sp(n)$ decomposes as the tensor
product of the fundamental representations of $Sp(1)$ and $O(n)$.
In $Sp(1)=SU(2)$, the equation $U_1U_2=-U_2U_1$ has a unique solution
up to conjugation. 
Let $U_1=A$, $U_2=B$ be such a solution.  Then in $Sp(n)$, we take
$U_1=A\otimes 1$, $U_2=B\otimes 1$, where $1$ is the identity element of 
$O(n)$.

The unbroken subgroup is $H=O(n)$, which has two components consisting
of elements of determinant 1 or $-1$.  The moduli space $\N$ of commuting
triples has two components, depending on whether $U_3$ is in the
identity or non-identity component of $H$.  If we set $U_3=1$, we
get the identity component $\N_0$ of $\N$.  The unbroken group is $SO(n)$,
of rank $r_0=n/2$ or $(n-1)/2$ for $n$ even or odd.  Setting $U_3={\rm diag}
(-1,1,1,\dots,1)$ we get a second component $\N_1$, with unbroken
$SO(n-1)$ and rank $r_1=n/2-1$ or $(n-1)/2$ for even or odd $n$.
Either way, the relation $(r_0+1)+(r_1+1)=h$ is obeyed, with $h=n+1$.

The nontrivial element of the center of $Sp(n)$ is the element $-1$,
which is in the identity or non-identity component of $H$ for even or
odd $n$.  Hence, in accord with what we found in section 3.4, there
is oblique confinement precisely if $n$ is odd.  For odd $n$, the
two components of $\N$ have the same rank and are permuted by the
action of the group $\Gamma$ of discrete gauge transformations, so that
zero energy states can carry nonzero electric flux.  For even $n$,
the two components of $\N$ have different rank and are each mapped
to themselves by $\Gamma$, so that zero energy states have vanishing
electric flux.

\bigskip\noindent{$Spin(2k+1)$}

The center of $Spin(2k+1)$ is again $\Z_2$.  A flat $Spin(2k+1)/\Z_2
=SO(2k+1)$ bundle with $m_{12}$ nonzero (and other components
of the magnetic flux vanishing) has holonomies $U_i$ that commute
in $SO(2k+1)$, but such that upon lifting to $Spin(2k+1)$, precisely
the same relations as \icono\ are obeyed.  

We can take $U_1={\rm diag}(-1,-1,1,\dots,1)$,
$U_2={\rm diag}(-1,1,-1,\dots,1)$ (each with precisely two
$-1$'s).  The unbroken subgroup of
$SO(2k+1)$ is $H=O(2k-2)$, with two components.  $U_3$ may be placed
in either component of $H$, giving representatives of the two components
of $\N$.  For one component $\N_0$, we can pick $U_3=1$, and the unbroken
group has connected component $SO(2k-2)$; for the  
second component $\N_1$, we can pick $U_3={\rm diag}(-1,-1,-1,-1,1,\dots,1)$
(with precisely four $-1$'s, ensuring that $U_3$ commutes with
$U_1,U_2$ even when lifted to $Spin(2k+1)$); the unbroken group has connected
component $SO(2k-3)$.  The familiar relation $(r_0+1)+(r_1+1)=h$ is obeyed,
with $h=2k-1$.

The nontrivial central element of $Spin(2k+1)$ is a $2\pi$ rotation,
which can be carried out in $H$ and is connected to the identity in $H$,
for all $k$.  Hence, as claimed in section
3.4,  there is no oblique confinement for $Spin(2k+1)$.  
The group $\Gamma$ maps each component of $\N$ to itself, and the
zero energy states have vanishing electric flux.

\bigskip\noindent{$Spin(4n+2)$}

The center of $Spin(4n+2)$ is $\Z_4$.  $Spin(4n+2)$ has an outer
automorphism that exchanges the two spinor representations and acts
on $m$ by $m\to -m$.  Modulo this automorphism, we may assume
that $m_{12}$, if nonzero, equals $1$ or $2$ mod 4.  For
$m_{12}=2$, the moduli space has two components that are
constructed just as for  $Spin(2k+1)$, with the same $U_1$ and $U_2$
and the same two choices for $U_3$.  There is one novelty compared 
to $Spin(2k+1)$: the center is $\Z_4$ instead of $\Z_2$, so the analysis
of oblique confinement is more elaborate.  The generator
of $\Z_4$ is an element $a$ that acts as $-1$ in the vector representation
and $\pm i$ on the two spinor representations.  $a$ is not contained in 
the connected component of the unbroken group $H$, but $a^2$ (which is
a $2\pi$ rotation in $SO(4n+2)$) is.
So  $w(G,m)=2$ for $m=2$.  The spectral flow character is of order 2.

The more novel case is $m_{12}=1$.
A flat $Spin(4n+2)$ bundle with $m_{12}=1$ can be constructed
as in section 3.4 using a subgroup $Spin(6)\times Spin(4)\times \dots
\times Spin(4)$ of $Spin(4n+2)$, with $n-1$ factors of $Spin(4)$.
In $Spin(6)=SU(4)$, we take $U_1U_2=iU_2U_1$, and in each factor
of $Spin(4)=SU(2)\times SU(2)$ we take $U_1=A\times 1$, $U_2=B\times 1$,
where $AB=-BA$.  The unbroken subgroup $H$ has $Sp(n-1)$, of rank $n-1$,
 for its
connected component.  This connected component does not contain
any element of the center of $Spin(4n+2)$.  The 
moduli space $\N$ of commuting triples has four components, one for
each element of the center of $Spin(4n+2)$; each component has a representative
in which $U_1$ and $U_2$ are as above
and $U_3$ is an element of the center.  The relation
$\sum_i(r_i+1)=h$ is obeyed with $h=4n$ and $r_i=n-1$ for
$i=0,\dots,3$.

\bigskip\noindent{$Spin(4n)$}

The center of $Spin(4n)$ is $\Z_2\times\Z_2$, with generators $a_1,a_2$
where $a_1$ is $1$ on positive chirality spinors, $-1$ on negative
chirality spinors, and $-1$ on vectors, while $a_2$ acts on those
representations as $-1,1,-1$.  For $m=a_1+a_2$, the construction
of the components of $\N$ is the same as for $Spin(2k+1)$.  The
subgroup of the center of $Spin(4n)$ that is contained in the connected
component of $H$ is  generated by $a_1+a_2$, so
as for $Spin(4n+2)$,
 $w(G,m)=2$ and there are zero energy states carrying electric flux.

Assuming that $m_{12}$ is the only nonzero component, the remaining
case (modulo an outer automorphism that exchanges $a_1$  and $a_2$)
is $m=a_1$.  To analyze this case, we use the embedding $O(2)\times O(2n)
\subset SO(4n)$, with the vector representation decomposing under the
subgroup as ${\bf 2}\otimes {\bf 2n}$.  We set 
\eqn\unicn{A=\left(\matrix{1 & 0 \cr 0 & -1}\right),~~~
B=\left(\matrix{0 & 1\cr 1 & 0 \cr}\right),}
and $U_1=A\times 1,$ $U_2=B\times 1$.  The unbroken subgroup $H$ is
a spin (or ``pin'')
double cover of $O(2n)$; the subgroup of the center of
$Spin(4n)$ contained in the identity component
of $H$ is $\Z_2$, generated by $a_1$ or $a_2$
depending on whether $n$ is even or odd.  So again $w(G,m)=2$ and
there are zero energy states carrying electric flux.  
The moduli space $\N$ of commuting triples has four components,
two of rank $n$ and two of rank $n-1$, leading to the formula
$\sum_i(r_1+1)=h$ with $h=4n-2$.  Two components contain  representatives
in which  $U_3$, if projected to $SO(4n)$, equals 1 (the actual $U_3$'s
differ in $Spin(4n)$ by a central element that projects to 1 in $SO(4n)$),
and two contain representatives in which $U$ projects to the element
${\rm diag}(-1,1,1,\dots,1)\in O(2n)\subset SO(4n)$.
In each case, $U_1$ and $U_2$ are as above.

\bigskip
This work was supported in part by NSF Grant PHY-9513835 and the Caltech
Discovery Fund.  I would
like to thank A. Borel, E. Diaconescu, G. Moore,
J. Morgan, N. Seiberg, and S. Weinberg
for helpful discussions.  Much of this material was presented at a
workshop at the ITP in Santa Barbara in January, 1998.  I would like
to thank the ITP for hospitality and the participants in the workshop
for questions and comments.
\listrefs
\end